\documentclass[12pt]{article}

\usepackage{cite} 
\usepackage{url}  
\usepackage{ifthen}  
\usepackage{multicol}   
\usepackage{multirow}
\usepackage{hhline}
\usepackage[utf8]{inputenc} 
\usepackage{fixltx2e} 
\usepackage{fullpage}
\usepackage{graphicx}
\usepackage{float}
\usepackage{color}
\usepackage{amsmath}
\usepackage{amsthm}
\usepackage{amssymb}
\usepackage{multirow}
\usepackage{rotating}
\usepackage[top=1.2in, bottom=1.5in, left=1in, right=1in]{geometry}
\usepackage[round]{natbib}

\usepackage{natbib}
\usepackage{apalike}

\usepackage{epsfig}
\usepackage{booktabs}

\usepackage{hyperref} 

\usepackage[usenames,dvipsnames,svgnames,table]{xcolor}

\usepackage{enumitem}

\usepackage{algorithm}
\usepackage{algpseudocode}

\usepackage{mathtools}
\DeclarePairedDelimiter{\ceil}{\lceil}{\rceil}

\usepackage[nodisplayskipstretch]{setspace}

\allowdisplaybreaks

\renewcommand{\cite}{\citep}

\newtheorem{theorem}{Theorem}
\newtheorem{lemma}{Lemma}
\newtheorem{proposition}{Proposition}

\newcommand{\beginsupplement}{%
        \setcounter{table}{0}
        \renewcommand{\thetable}{S\arabic{table}}%
        \setcounter{figure}{0}
        \renewcommand{\thefigure}{S\arabic{figure}}%
     }
     
\makeatletter
\let\@fnsymbol\@arabic
\makeatother

\newcommand{\bbX}{\mathbb{X}}
\newcommand{\bbR}{\mathbb{R}}

\newcommand{\bx}{\boldsymbol{x}}
\newcommand{\bX}{\boldsymbol{X}}

\newcommand{\boldmu}{\boldsymbol{\mu}}
\newcommand{\boldSigma}{\boldsymbol{\Sigma}}
\newcommand{\sigb}{\sigma_b}

\title{Statistical Significance for Hierarchical Clustering}

\author{%
Patrick~K.~Kimes$^1$ \and %
Yufeng~Liu$^{1,2,3}$ \and %
D.~Neil~Hayes$^4$ \and %
J.~S.~Marron$^{1,2}$}

\date{}

\begin{document}

\maketitle

\let\thefootnote\relax\footnote{\noindent
$^1$Department of Statistics and Operations Research, 
$^2$Department of Biostatistics, 
$^3$Carolina Center for Genome Sciences, 
$^4$Lineberger Comprehensive Cancer Center, University of North Carolina, Chapel Hill, NC 27599.
}



\noindent
{\bf Abstract:}
Cluster analysis has proved to be an invaluable tool for the exploratory and unsupervised analysis of high dimensional datasets. Among methods for clustering, hierarchical approaches have enjoyed substantial popularity in genomics and other fields for their ability to simultaneously uncover multiple layers of clustering structure. A critical and challenging question in cluster analysis is whether the identified clusters represent important underlying structure or are artifacts of natural sampling variation. Few approaches have been proposed for addressing this problem in the context of hierarchical clustering, for which the problem is further complicated by the natural tree structure of the partition, and the multiplicity of tests required to parse the layers of nested clusters. In this paper, we propose a Monte Carlo based approach for testing statistical significance in hierarchical clustering which addresses these issues. The approach is implemented as a sequential testing procedure guaranteeing control of the family-wise error rate. Theoretical justification is provided for our approach, and its power to detect true clustering structure is illustrated through several simulation studies and applications to two cancer gene expression datasets.
\vskip.25in


\thispagestyle{empty}
\clearpage
\setcounter{page}{1}


\section{INTRODUCTION}
\label{introduction}

Clustering describes the unsupervised learning task of partitioning observations into homogenous subsets, or clusters, to uncover subpopulation structure in a dataset. As an unsupervised learning task, cluster analysis makes no use of label or outcome data. A large number of methods have been proposed for clustering, including hierarchical approaches, as well as non-nested approaches, such as $K$-means clustering. Since the work of \citet{Eisen1998}, hierarchical clustering algorithms have enjoyed substantial popularity for the exploratory analysis of gene expression data. In several landmark papers that followed, these methods were successfully used to identify clinically relevant expression subtypes in lymphoma, breast, and other types of cancer \cite{Alizadeh2000,Perou2000,Bhattacharjee2001}.

While non-nested clustering algorithms typically require pre-specifying the number of clusters of interest, $K$, hierarchical algorithms do not. Instead, hierarchical approaches produce a single nested hierarchy of clusters from which a partition can be obtained for any feasible $K$. As a result, hierarchical clustering provides an intuitive way to study relationships among clusters not possible using non-nested approaches. The popularity of hierarchical clustering in practice may also be largely attributed to \emph{dendrograms}, a highly informative visualization of the clustering as a binary tree.

While dendrograms provide an intuitive representation for studying the results of hierarchical clustering, the researcher is still ultimately left to decide which partitions along the tree to interpret as biologically important subpopulation differences. Often, in genomic studies, the determination and assessment of subpopulations are left to heuristic or \textit{ad hoc} methods \cite{Verhaak2010,Wilkerson2010,Bastien2012}. To provide a statistically sound alternative to these methods, we introduce statistical Significance of Hierarchical Clustering (SHC), a Monte Carlo based approach for assessing the statistical significance of clustering along a hierarchical partition. The approach makes use of the ordered and nested structure in the output of hierarchical clustering to reduce the problem to a sequence of hypothesis tests. Each test is formulated such that the procedure may be applied even in the high-dimension low-sample size (HDLSS) setting, where the number of variables is much greater than the number of observations. This is of particular importance, as the number of measurements being made in genomic studies continues to grow with advances in high-throughput sequencing technologies, such as RNA-seq \cite{Marioni2008,Wang2009a}. A stopping rule along the sequence of tests is also provided to control the family-wise error rate (FWER) of the entire procedure.

Several approaches have been proposed to address the question of statistical significance in the non-nested setting. The Statistical Significance of Clustering (SigClust) hypothesis test was introduced by \citet{Liu2008} for assessing the significance of clustering in HDLSS settings using a Monte Carlo procedure. While well-suited for detecting the presence of more than a single cluster in a dataset, the approach was not developed with the intention of testing in hierarchical or multi-cluster settings. This approach is described in greater detail in Section~\ref{background:sigclust}. More recently, \citet{Maitra2012} proposed a bootstrap based approach capable of testing for any number of clusters in a dataset. However, in addition to not directly addressing the hierarchical problem, their approach has not been evaluated in the important HDLSS setting. As such, neither approach provides a solution for handling the structure and multiplicity of nested tests unique to hierarchical clustering. 

For assessing statistical significance in the hierarchical setting, \citet{Suzuki2006} developed the \texttt{R} package \texttt{pvclust}. The hypothesis tests used in \texttt{pvclust} are based on bootstrapping procedures originally proposed for significance testing in the context of phylogenetic tree estimation \cite{Efron1996,Shimodaira2004}. Since the procedure is based on a nonparamateric bootstrapping of the covariates, while \texttt{pvclust} can be used in the HDLSS setting, it cannot be implemented when the dataset is of low-dimension. In contrast, SHC may be used in either setting. The overall approach of \texttt{pvclust} differs fundamentally from that of SHC and is discussed briefly in Section~\ref{simulations}. To our knowledge, no other approaches have been proposed for assessing the statistical significance of hierarchical clustering. 

The remainder of this paper is organized as follows. In Section~\ref{background} we first review hierarchical clustering and describe the SigClust hypothesis test of \citet{Liu2008}. Then, in Section~\ref{methodology}, we introduce our proposed SHC approach. In Section~\ref{theoretical}, we present theoretical justifications for our method under the HDLSS asymptotic setting. We then evaluate the performance of our method under various simulation settings in Section~\ref{simulations}. In Section~\ref{realdata}, we apply our method to two cancer gene expression datasets. Finally, we conclude with a discussion in Section~\ref{discussion}. The SHC procedure is implemented in \texttt{R}, and is available at the first author's website. 

\section{CLUSTERING AND SIGNIFICANCE}
\label{background}
We begin this section by first providing a brief review of hierarchical clustering. We then describe the $K$-means based SigClust approach of \citet{Liu2008} for assessing significance of clustering in HDLSS data.

\subsection{Hierarchical Clustering Methods}
Given a collection of $N$ unlabeled observations, $\mathbb{X} = \{\bx_i,\ldots, \bx_N\}$, algorithms for hierarchical clustering estimate all $K=1,\ldots,N$ partitions of the data through a sequential optimization procedure. The sequence of steps can be implemented as either an agglomerative (bottom-up) or divisive (top-down) approach to produce the nested hierarchy of clusters. Agglomerative clustering begins with each observation belonging to one of $N$ disjoint singleton clusters. Then, at each step, the two most similar clusters are joined until after $(N-1)$ steps, all observations belong to a single cluster of size $N$. Divisive clustering proceeds in a similar, but reversed manner. In this paper we focus on agglomerative approaches which are more often used in practice. 

Commonly, in agglomerative clustering, the pairwise similarity of observations is measured using a \emph{dissimilarity function}, such as squared Euclidean distance ($L_2^2$), Manhattan distance ($L_1$), or $(1 - |\text{Pearson corr.}|)$. Then, a \emph{linkage function} is used to extend this notion of dissimilarity to pairs of clusters. Often, the linkage function is defined with respect to all pairwise dissimilarities of observations belong to the separate clusters. Examples of linkage functions include Ward's, single, complete, and average linkage \cite{Ward1963}. The clusters identified using hierarchical algorithms depend heavily on the choice of both the dissimilarity and linkage functions.

\begin{figure}[t!]
	\centering
	\includegraphics[width=0.7\textwidth]{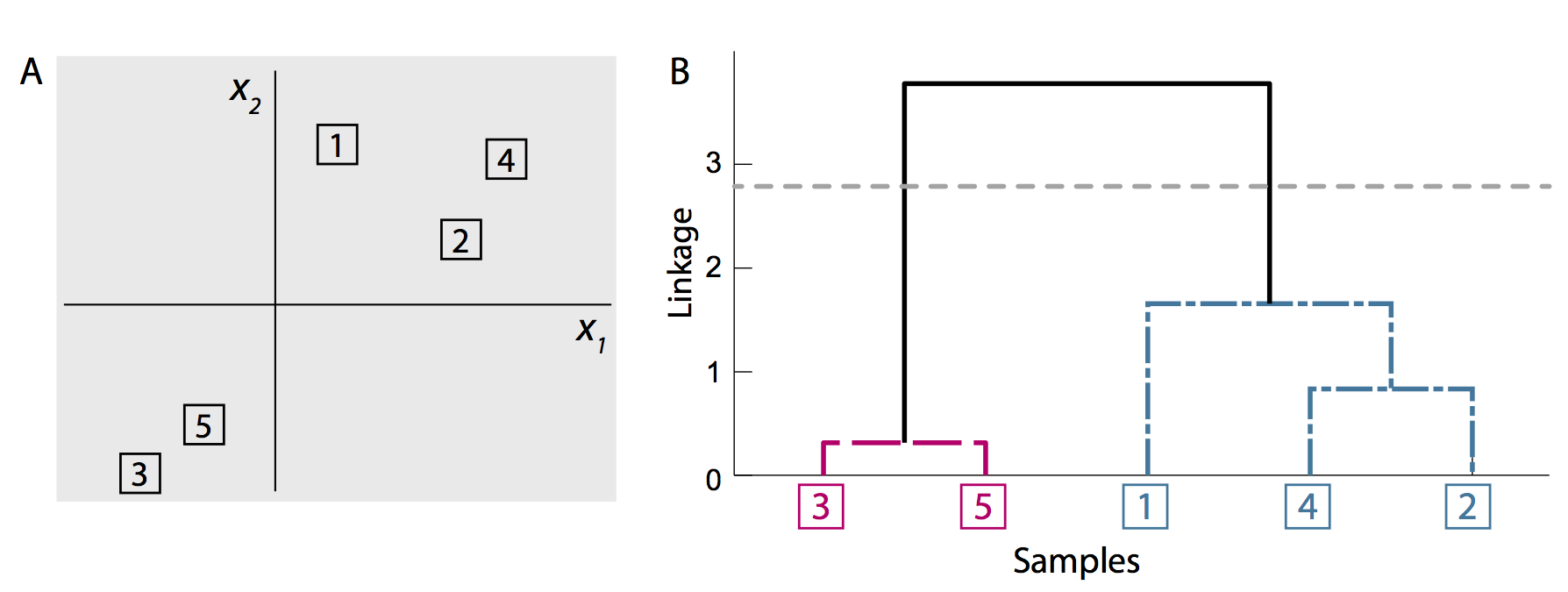}
	\caption[Example of hierarchical clustering and a dendrogram.]{Hierarchical clustering applied to 5 observations. (A) Scatterplot of the observations. (B) The corresponding dendrogram.}
	\label{fig:dendrogram}
\end{figure}

The sequence of clustering solutions obtained by hierarchical clustering is naturally visualized as a binary tree, commonly referred to as a dendrogram. Figure~\ref{fig:dendrogram}A shows a simple example with five points in $\bbR^2$ clustered using squared Euclidean dissimilarity and average linkage. The corresponding dendrogram is shown in Figure~\ref{fig:dendrogram}B, with the observation indices placed along the horizontal axis, such that no two branches of the dendrogram cross. The sequential clustering procedure is shown by the joining of clusters at their respective linkage value, denoted by the vertical axis of Figure~\ref{fig:dendrogram}B, such that the most similar clusters and observations are connected near the bottom of the tree. The spectrum of clustering solutions can be recovered from the dendrogram by cutting the tree at an appropriate height, and taking the resulting subtrees as the clustering solution. For example, the corresponding $K=2$ solution is obtained by cutting the dendrogram at the gray horizontal line in Figure~\ref{fig:dendrogram}B.

\subsection{Statistical Significance}
\label{background:sigclust}
We next describe the SigClust hypothesis test of \citet{Liu2008} for assessing significance of clustering. To make inference in the HDLSS setting possible, SigClust makes the simplifying assumption that a cluster may be characterized as a subset of the data which follows a single Gaussian distribution. Therefore, to determine whether a dataset is comprised of more than a single cluster, the approach tests the following hypotheses:
\begin{align*}
	H_0:&\ \text{the data follow a single Gaussian distribution} && \\ 
	H_1:&\ \text{the data follow a non-Gaussian distribution.} &&
\end{align*}
The corresponding $p$-value is calculated using the 2-means cluster index (CI), a statistic sensitive to the null and alternative hypotheses. Letting $C_k$ denote the set of indices of observations in cluster $k$ and using $\bar\bx_k$ to denote the corresponding cluster mean, the 2-means CI is defined as
\begin{align}
	\text{CI} &
		= \frac{\sum_{k=1}^2 \sum_{i\in C_k} \| \bx_i - \bar \bx_k \|_2^2}{\sum_{i=1}^N \| \bx_i - \bar \bx \|_2^2}
		= \frac{SS_1 + SS_2}{TSS}, \label{eq:2CI}
\end{align}
where $TSS$ and $SS_k$ are the total and within-cluster sum of squares. Smaller values of the 2-means CI correspond to tighter clusters, and provide stronger evidence of clustering of the data. The statistical significance of a given pair of clusters is calculated by comparing the observed 2-means CI against the distribution of 2-means CIs under the null hypothesis of a single Gaussian distribution. Since a closed form of the distribution of CIs under the null is unavailable, it is empirically approximated by the CIs computed for hundreds, or thousands, of datasets simulated from a null Gaussian distribution estimated using the original dataset. An empirical $p$-value is calculated by the proportion of simulated null CIs less than the observed CI. Approximations to the optimal 2-means CI for both the observed and simulated datasets can be obtained using the $K$-means algorithm for two clusters.

In the presence of strong clustering, the empirical $p$-value may simply return $0$ if all simulated CIs fall above the observed value. This can be particularly uninformative when trying to compare the significance of multiple clustering events. To handle this problem, \citet{Liu2008} proposed computing a \emph{Gaussian fit $p$-value} in addition to the empirical $p$-value. Based on the observation that the distribution of CIs appears roughly Gaussian, the Gaussian fit $p$-value is calculated as the lower tail probability of the best-fit Gaussian distribution to the simulated null CIs.

An important issue not discussed above is the estimation of the covariance matrix of the null distribution, a non-trivial task in the HDLSS setting. A key part of the SigClust approach is the simplification of this problem, by making use of the invariance of the 2-means CI to translations and rotations of the data in the Euclidean space. It therefore suffices to simulate data from an estimate of any rotation and shift of the null distribution. Conveniently, by centering the distribution at the origin, and rotating along the eigendirections of the covariance matrix, the task can be reduced to estimating only the eigenvalues of the null covariance matrix. As a result, the number of parameters to estimate is reduced from $p(p+1)/2$ to $p$. However, in the HDLSS setting, even the estimation of $p$ parameters is challenging, as $N \ll p$. To solve this problem, the additional assumption is made that the null covariance matrix follows a factor analysis model. That is, under the null hypothesis, the observations are assumed to be drawn from a single Gaussian distribution, ${N}(\boldsymbol\mu,\Sigma)$, with $\Sigma$ having eigendecomposition $\Sigma = U\Lambda U^T$ such that 
\begin{align*}
	\Lambda = \Lambda_0 + \sigma_b^2 \textbf{I}_p,
\end{align*}
where $\Lambda_0$ is a low rank ($<N$) diagonal matrix of true signal, $\sigma^2_b$ is a relatively small amount of background noise, and $\textbf{I}_p$ is the $p$-dimensional identity matrix. Letting $w$ denote the number of non-zero entries of $\Lambda_0$, under the factor analysis model, only $w+1$ parameters must be estimated to implement SigClust. Several approaches have been proposed for estimating $\sigb^2$ and the $w$ non-zero entries of $\Lambda_0$, including the hard-threshold, soft-threshold, and sample-based approaches \cite{Liu2008,Huang2013}.

\section{METHODOLOGY}
\label{methodology}
To assess significance of clustering in a hierarchical partition, we propose a sequential testing procedure in which Monte Carlo based hypothesis tests are preformed at select nodes along the corresponding dendrogram. In this section, we introduce our SHC algorithm in two parts. First, using a toy example, we describe the hypothesis test performed at individual nodes. Then, we describe our sequential testing procedure for controlling the FWER of the algorithm along the entire dendrogram. 

\subsection{SHC Hypothesis Test}
\begin{figure}[t!]
	\centering
	\includegraphics[width=.9\textwidth]{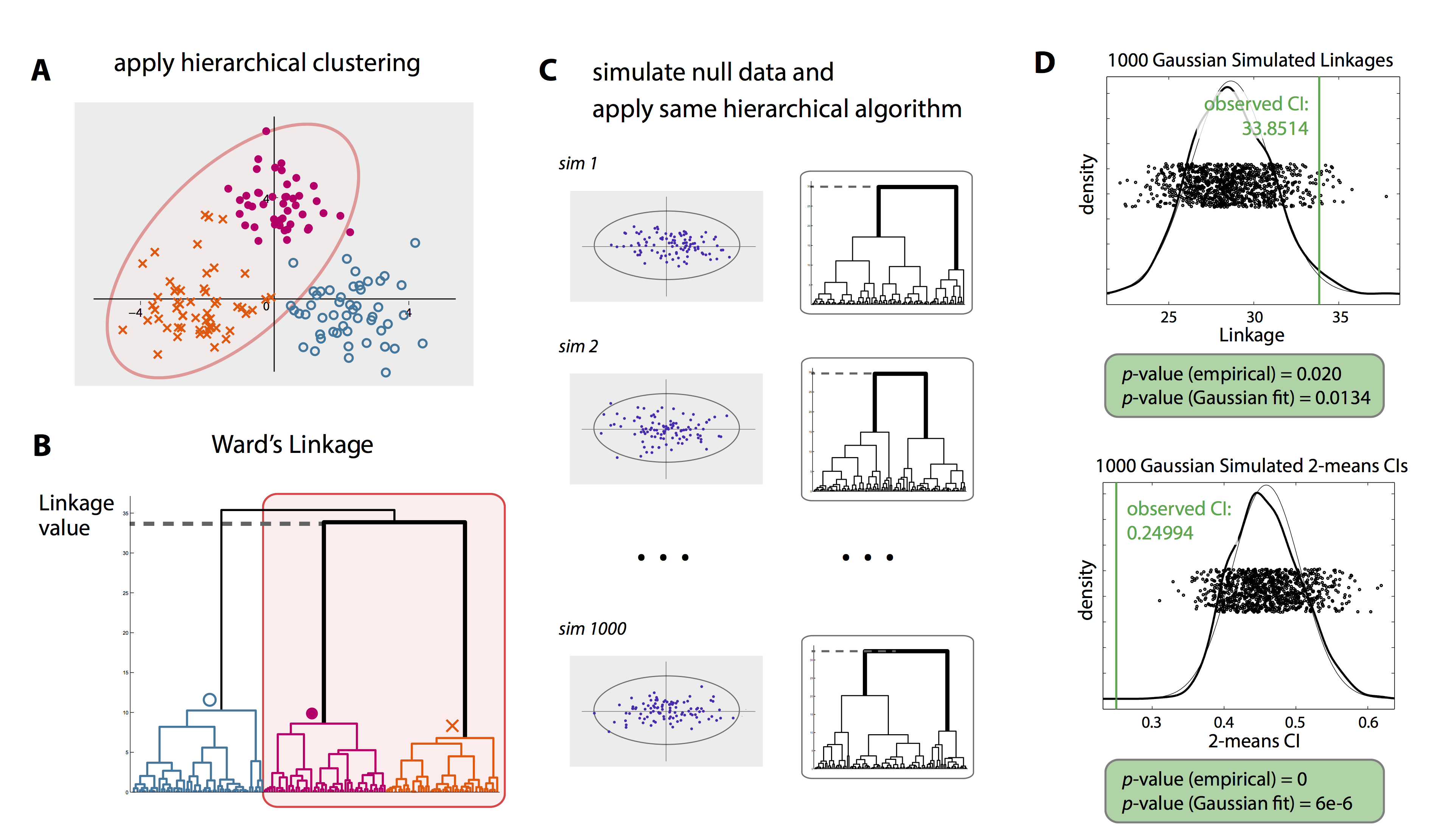}%
	\caption[Hierarchical SigClust workflow]{The SHC testing procedure illustrated using a toy example. Testing is applied to the 96 observations joined at the second node from the root. (A) Scatterplot of the observations in $\mathbb{R}^2$. (B) The corresponding dendrogram. (C) Hierarchical clustering applied to 1000 datasets simulated from a null Gaussian estimated from the 96 observations. (D) Distributions of null cluster indices used to calculate the empirical SHC $p$-values.}
	\label{fig:hierarchicalworkflow}
\end{figure}

Throughout, we use $j \in \{1,\ldots,N-1\}$ to denote the node index, such that $j=1$ and $j=(N-1)$ correspond to the first (lowest) and final (highest) merges along the dendrogram. In Figure~\ref{fig:hierarchicalworkflow}, we illustrate one step of our sequential algorithm using a toy dataset of $N=150$ observations drawn from $\mathbb{R}^2$ (Figure~\ref{fig:hierarchicalworkflow}A). Agglomerative hierarchical clustering was applied using Ward's linkage to obtain the dendrogram in Figure~\ref{fig:hierarchicalworkflow}B. Consider the second node from the top, i.e. $j=(N-2)$. The corresponding observations and subtree are highlighted in panels A and B of Figure~\ref{fig:hierarchicalworkflow}. Here, we are interested in whether the sets of 43 and 53 observations joined at node $(N-2)$, denoted by dots and $\times$'s, more naturally define one or two distinct clusters. Assuming that a cluster may be well approximated by a single Gaussian distribution, we propose to test the following hypotheses at node $(N-2)$:
\begin{align*}
	H_0:&\ \text{The 96 observations follow a single Gaussian distribution} && \\ 
	H_1:&\ \text{The 96 observations do not follow a single Gaussian distribution.} &&
\end{align*}
The $p$-value at the node, denoted by $p_j$, is calculated by comparing the strength of clustering in the observed data against that for data clustered using the same hierarchical algorithm under the null hypothesis. We consider two cluster indices, linkage value and the 2-means CI, as natural measures for the strength of clustering in the hierarchical setting. To approximate the null distribution of cluster indices, 1000 datasets of 96 observations are first simulated from a null Gaussian distribution estimated using only the 96 observations included in the highlighted subtree. Then, each simulated dataset is clustered using the same hierarchical algorithm as was applied to the original dataset (Figure~\ref{fig:hierarchicalworkflow}C). As with the observed data, the cluster indices are computed for each simulated dataset using the two cluster solution obtained from the hierarchical algorithm. Finally, $p$-values are obtained from the proportion of null cluster indices indicating stronger clustering than the observed indices (Figure~\ref{fig:hierarchicalworkflow}D). For the linkage value and 2-means CI, this corresponds to larger and smaller values, respectively. As in SigClust, we also compute a Gaussian approximate $p$-value in addition to the empirical $p$-value. In this example, the resulting empirical $p$-values, $0.020$ and $0$, using linkage and the 2-means CI, both suggest significant clustering at the node.
  
In estimating the null Gaussian distribution, we first note that many popular linkage functions, including Ward's, single, complete and average, are defined with respect to the pairwise dissimilarities of observations belonging to two clusters. As such, the use of these linkage functions with any dissimilarity satisfying translation and rotation invariance, such as Euclidean or squared Euclidean distance, naturally leads to the invariance of the entire hierarchical procedure. Thus, for several choices of linkage and dissimilarity, the SHC $p$-value can be equivalently calculated using data simulated from a simplified distribution centered at the origin, with diagonal covariance structure. To handle the HDLSS setting, as in SigClust, we further assume that the covariance matrix of the null Gaussian distribution follows a factor analysis model, such that the problem may be addressed using the hard-threshold, soft-threshold and sample approaches previously proposed in \citet{Liu2008,Huang2013}. 

Throughout this paper we derive theoretical and simulation results using squared Euclidean dissimilarity with Ward's linkage, an example of a translation and rotation invariant choice of dissimilarity and linkage function. However, our approach may be implemented using a larger class of linkages and appropriately chosen dissimilarity functions. We focus on Ward's linkage clustering as the approach may be interpreted as characterizing clusters as single Gaussian distributions, as in the hypotheses we propose to test. Additionally, we have observed that Ward's linkage clustering often provides strong clustering results in practice. 

Note that at each node, the procedure requires fitting a null Gaussian distribution using only the observations contained in the corresponding subtree. We therefore set a minimum subtree size, $N_{\text{min}}$, for testing at any node. For the simulations described in Section~\ref{simulations}, we use $N_{\text{min}}=10$.

In this section, we have described only a single test of the entire SHC procedure. For a dataset of $N$ observations, at most $(N-1)$ tests may be performed at the nodes along the dendrogram. While the total possible number of tests is typically much smaller due to the minimum subtree criterion, care is still needed to account for the issue of multiple testing. In the following section, we describe a sequential approach for controlling the FWER to address this issue.

\subsection{Multiple Testing Correction}
\label{methodology:FWER}
To control the FWER of the SHC procedure, one could simply test at all nodes simultaneously, and apply an equal Bonferroni correction to each test. However, this approach ignores the clear hierarchical nature of the tests. Furthermore, the resulting dendrogram may have significant calls at distant and isolated nodes, making the final output difficult to interpret. Instead, we propose to control the FWER using a sequential approach which provides greater power at the more central nodes near the root of the dendrogram, and also leads to more easily interpretable results.

To correct for multiple testing, we employ the FWER controlling procedure of \citet{Meinshausen2008} original proposed in the context of variable selection. For the SHC approach, the FWER along the entire dendrogram is defined to be the probability of at least once, falsely rejecting the null at a subtree of the dendrogram corresponding to a single Gaussian cluster. To control the FWER at level $\alpha \in (0,1)$, we perform the hypothesis test described above at each node $j$, with the modified significance cutoff: 
\begin{align*}
	\alpha^*_j = \alpha \cdot \frac{N_j-1}{N-1},
\end{align*}
where $N_j$ is used to denote the number of observations clustered at node $j$. Starting from the root node, i.e. $j = (N-1)$, we descend the dendrogram rejecting at nodes for which the following two conditions are satisfied: (C1) $p_j < \alpha^*_j$, and (C2) the parent node was also rejected, where the parent of a node is simply the one directly above it. For the root node, condition (C2) is ignored. As the procedure moves down the dendrogram, condition (C1) and the modified cutoff, $\alpha^*_j$, apply an increasingly stringent correction to each test, proportional to the size of the corresponding subtree. Intuitively, if the subtree at a node contains multiple clusters, the same is true of any node directly above it. Condition (C2) formalized this intuition by forcing the set of significant nodes to be well connected from the root. Furthermore, recall that the hypotheses tested at each node assess whether or not the two subtrees were generated from a single Gaussian distribution. While appropriate when testing at nodes which correspond to one or more Gaussian distributions, the interpretation of the test becomes more difficult when applied to only a portion of a single Gaussian distribution, e.g. only half of a Gaussian cluster. This can occur when testing at a node which falls below a truly null node. In this case, while the two subtrees of the node correspond to non-Gaussian distributions, they do not correspond to interesting clustering behavior. Thus, testing at such nodes may result in truly positive, but uninteresting, significant calls. By restricting the set of significant nodes to be well connected from the root, in addition to controlling the FWER, our procedure also limits the impact of such undesirable tests.

\section{THEORETICAL DEVELOPMENT}
\label{theoretical}

In this section, we study the theoretical behavior of our SHC procedure with linkage value as the measure of cluster strength applied to Ward's linkage hierarchical clustering. We derive theoretical results for the approach under both the null and alternative hypotheses. In the null setting, the data are sampled from a single Gaussian distribution. Under this setting, we show that the empirical SHC $p$-value at the root node follows the $U(0,1)$ distribution. In the alternative setting, we consider the case when the data follow a mixture of two spherical Gaussian distributions. Since SHC is a procedure for assessing statistical significance given a hierarchical partition, the approach depends heavily on the algorithm used for clustering. We therefore first provide conditions for which Ward's linkage clustering asymptotically separates samples from the two components at the root node. Given these conditions are satisfied, we then show that the corresponding empirical SHC $p$-value at the root node tends to 0 asymptotically as both the sample size and dimension grow to infinity. All proofs are included in the Appendix.

We first consider the null case where the data, $\bbX = \{\bX_1,\ldots,\bX_N\}$, are sampled from a single Gaussian distribution, $N(\bold{0}, \boldSigma)$. The following proposition describes the behavior of the empirical $p$-value at the root node under this setting.

\begin{proposition}\label{prop:level}
	Suppose $\bbX$ were drawn from a single Gaussian distribution, $N(\bold{0}, \boldSigma)$, with known covariance matrix $\boldSigma$. Then, the SHC empirical $p$-value at the root node follows the $\bold{U}(0,1)$ distribution.
\end{proposition}

The proof of Proposition~\ref{prop:level} is omitted, as it follows directly from an application of the probability integral transform. We also note that the result of Proposition~\ref{prop:level} similarly holds for any subtree along a dendrogram corresponding to a single Gaussian distribution. Combining this with Theorem~1 of \citet{Meinshausen2008}, we have that the modified $p$-value cutoff procedure of Section~\ref{methodology:FWER} controls the FWER at the desired level $\alpha$. 

We next consider the alternative setting. Suppose the data, $\bbX$, were drawn from a mixture of two Gaussian subpopulations in $\mathbb{R}^p$, denoted by $N(\boldmu_1, \sigma_1^2 \textbf{I}_p)$ and $N(\boldmu_2, \sigma_2^2 \textbf{I}_p)$. Let $\bbX^{(1)} = \{\bX^{(1)}_1, \ldots, \bX^{(1)}_n\}$ and $\mathbb{X}^{(2)} = \{\bX^{(2)}_1, \ldots, \bX^{(2)}_m\}$ denote the $N=n+m$ observations of $\bbX$ drawn from the two mixture components. In the following results, we consider the HDLSS asymptotic setting where $p\rightarrow\infty$ and $n = p^\alpha + o(p)$, $m = p^\beta + o(p)$ for $\alpha, \beta \in (0,1)$ \cite{Hall2005}. As in \citet{Borysov2014}, we assume that the mean of the difference $(\bX^{(1)}_i - \bX^{(2)}_j)$ is not dominated by a few large coordinates in the sense that for some $\epsilon>0$,
\begin{align}
	\sum_{k=1}^p \left( \mu_{1,k} - \mu_{2, k}\right)^4 = o\left(p^{2-\epsilon}\right), \ \ \ \ p\rightarrow\infty. \label{eq:L4bound}
\end{align}
Given this assumption, the following theorem provides necessary conditions for Ward's linkage clustering to correctly separate observations of the two mixture components.

\begin{theorem}\label{thm:wardsplit}
	Suppose (\ref{eq:L4bound}) is satisfied and the dendrogram is constructed using the Ward's linkage function. Let $n,m$ be the number of observations sampled from the two Gaussian mixture components, $N(\boldmu_1, \sigma_1^2 \textbf{I}_p)$ and $N(\boldmu_2, \sigma_2^2 \textbf{I}_p)$, with $\sigma_1 \leq \sigma_2$. Additionally, suppose $n = p^\alpha + o(p)$, $m = p^\beta + o(p)$ for $\alpha, \beta \in (0,1)$, and let $\mu^2$ denote $p^{-1}\| \boldmu_1 - \boldmu_2 \|_2^2$. Then, if $\lim\sup \frac{n^{-1} (\sigma_2^2 - \sigma_1^2)}{\mu^2} < 1$, $\bX^{(1)}$ and $\bX^{(2)}$ are separated at the root node with probability converging to 1 as $p\rightarrow\infty$.
\end{theorem}

Theorem~\ref{thm:wardsplit} builds on the asymptotic results for hierarchical clustering described in \citet{Borysov2014}. The result provides a theoretical analysis of Ward's linkage clustering, independent of our SHC approach. In the following result, using Theorem~\ref{thm:wardsplit}, we show that under further assumptions, the SHC empirical $p$-value is asymptotically powerful at the root node of the dendrogram. That is, the $p$-value converges to $0$ as $p,n,m$ grow to infinity.

\begin{theorem}\label{thm:wardpower}
	Suppose the assumptions for Theorem~\ref{thm:wardsplit} are satisfied. Furthermore, suppose $\sigma_1^2$ and $\sigma_2^2$ are known. Then, using linkage as the measure of cluster strength, the empirical SHC $p$-value at the root node along the dendrogram equals 0 with probability converging to 1 as $p\rightarrow\infty$.
\end{theorem}

By Theorem~\ref{thm:wardpower}, the SHC procedure is asymptotically well powered to identify significant clustering structure in the presence of multiple Gaussian components. While in this section we only considered the theoretical behavior of SHC using linkage value as the measure of cluster strength, empirical results presented in the following section provide justification for alternatively using the 2-means CI. In the next section, we compare the power and level of SHC using linkage value and the 2-means CI and other approaches through several simulation studies.

\section{SIMULATIONS}
\label{simulations}
In this section we illustrate the performance of our proposed SHC approach using simulation studies. In Section~\ref{methodology}, we described SHC as the combination of two elements: (1) a sequential testing scheme for controlling the FWER applied to the results of hierarchical clustering, and (2) a simulation-based hypothesis test for assessing the statistical significance of hierarchical clustering at a single node. To evaluate the advantage of tuning the test at each node for hierarchical clustering, we consider two implementations of our SHC approach, denoted by SHC1 and SHC2. In SHC1, we combine our proposed iterative testing scheme with the classical SigClust test applied at each node. In SHC2, we implement our complete procedure, which directly accounts for the effect of hierarchical clustering in the calculation of the $p$-value. Two implementations of SHC2 are further considered, denoted by $\text{SHC2}_L$ and $\text{SHC2}_2$, differing by whether the linkage value or the 2-means CI is used to measure the strength of clustering. Note that both SHC1 and SHC2 may be viewed as contributions of our work with differing levels of adjustment for the hierarchical setting. 

The performance of SHC1 and SHC2 are compared against the existing \texttt{pvclust} approach. In each simulation, Ward's linkage clustering was applied to a dataset drawn from a mixture of Gaussian distributions in $\mathbb{R}^p$. A range of simulation settings were considered, including the null setting with $K=1$ and alternative settings with $K=2$, $K=3$, and $K=4$. A representative set of simulation results for $K=1$, $K=3$ and $K=4$ are reported in this section. As the $K=2$ setting reduces to a non-nested clustering problem, these results are omitted from the main text. However, complete simulation results, including the entire set of $K=2$ results (Supplementary Table~S2), may be found in the Supplementary Materials.

In all simulations, SHC1 and SHC2 $p$-values were calculated using 100 simulated null cluster indices, and the corresponding Gaussian-fit $p$-values are reported. When $p > n$, the covariance matrix for the Gaussian null was estimated using the soft-threshold approach described in \citet{Huang2013}. The \texttt{pvclust} approach was implemented using 1000 bootstrap samples, as suggested in \citet{Suzuki2006}. However, to keep the total computational time of the entire set of simulations manageable, the simulations reported in the Supplementary Materials were completed using 100 bootstrap samples. Results for \texttt{pvclust} are only reported for high dimensional simulations, as the approach does not appear to be able to handle datasets in lower dimensions, e.g. $p=2$. All simulation settings were replicated 100 times. Before presenting the simulation results, we first provide a brief review of the fundamental difference between \texttt{pvclust} and our proposed SHC method. 

The \verb+pvclust+ method of \citet{Suzuki2006} computes two values: an approximately unbiased (AU) $p$-value based on a multi-step multi-scale bootstrap resampling procedure \cite{Shimodaira2004}, and a bootstrap probability (BP) $p$-value calculated from ordinary bootstrap resampling \cite{Efron1996}. Similar to SHC, \texttt{pvclust} also tests at nodes along the dendrogram. However, no test is performed at the root node, and the corresponding hypotheses tested at each node is given by:
\begin{align*}
	H_0&: \text{ the cluster does not exist} \\
	H_1&: \text{ the cluster exists}.
\end{align*}
The difference between the two approaches can be understood by examining the dendrogram presented in Figure~\ref{fig:hierarchicalworkflow}B. Using SHC, significant evidence of the three clusters is obtained if the null hypothesis is rejected at the top two nodes of the dendrogram. In contrast, to identify the three clusters using \texttt{pvclust}, the null hypothesis must be rejected at the three nodes directly above each cluster, denoted by their respective cluster symbol.

\subsection{Null Setting ($K=1$)}
\label{sim:K1}

\begin{table}[t!]\footnotesize
	\centering
	
\begin{tabular}{ccc rrrrr rrrr}
\toprule
\multicolumn{3}{c}{parameters}&
	\multicolumn{5}{c}{$|p\text{-value}<0.05|$ (mean $p$-value)}&
	\multicolumn{4}{c}{median time (sec.)}
\tabularnewline
\cmidrule(r){1-3} \cmidrule(r){4-8} \cmidrule(r){9-12}
\multicolumn{1}{c}{$N$}&
	\multicolumn{1}{c}{$w$}&
	\multicolumn{1}{c}{$v$}&
	\multicolumn{1}{c}{\texttt{pv}AU}&
	\multicolumn{1}{c}{\texttt{pv}BP}&
	\multicolumn{1}{c}{SHC1}&
	\multicolumn{1}{c}{$\text{SHC2}_L$}&
	\multicolumn{1}{c}{$\text{SHC2}_2$}&
	\multicolumn{1}{c}{\texttt{pv}}&
	\multicolumn{1}{c}{SHC1}&
	\multicolumn{1}{c}{$\text{SHC2}_L$}&
	\multicolumn{1}{c}{$\text{SHC2}_2$}
\tabularnewline
\midrule
$ 50$ & $0$ & $-$  & $0\ (0.99)$ & $0\ (1.00)$ & $0\ (1.00)$ & $0\ (1.00)$ & $0\ (1.00)$ & $ 30.61^*$ & $20.56$ & $12.33$ & $14.59$ \tabularnewline
$ 50$ & $1$ & 100  & $2\ (0.48)$ & $0\ (0.98)$ & $0\ (0.59)$ & $0\ (0.49)$ & $0\ (0.47)$ & $ 30.54^*$ & $22.31$ & $13.35$ & $15.60$ \tabularnewline
$ 50$ & $5$ & 100  & $1\ (0.61)$ & $0\ (1.00)$ & $0\ (0.83)$ & $0\ (0.73)$ & $0\ (0.65)$ & $ 30.52^*$ & $21.11$ & $12.68$ & $14.82$ \tabularnewline
\addlinespace
$100$ & $0$ & $-$  & $0\ (1.00)$ & $0\ (1.00)$ & $0\ (1.00)$ & $0\ (1.00)$ & $0\ (1.00)$ & $108.52^*$ & $48.18$ & $29.19$ & $35.04$ \tabularnewline
$100$ & $1$ & 100  & $2\ (0.89)$ & $0\ (1.00)$ & $0\ (0.69)$ & $0\ (0.49)$ & $0\ (0.49)$ & $108.70^*$ & $50.49$ & $30.73$ & $36.85$ \tabularnewline
$100$ & $5$ & 100  & $1\ (0.98)$ & $0\ (1.00)$ & $0\ (0.96)$ & $0\ (0.72)$ & $0\ (0.72)$ & $108.85^*$ & $51.04$ & $30.74$ & $37.01$ \tabularnewline
\bottomrule
\end{tabular}

	\caption{Simulation~\ref{sim:K1} ($K=1$). Number of false positives at $\alpha=0.05$, mean $p$-value, median computation time over 100 replications. $*$: \texttt{pvclust} times scaled by $1/10$.}
	\label{table:K1results}
\end{table}

We first consider the null setting  to evaluate the ability of SHC to control for false positives. In these simulations, datasets of size $N=50$ and $100$ were sampled from a single Gaussian distribution in $p=1000$ dimensions with diagonal covariance structure given by:
\begin{align*}
	\boldSigma = \text{diag}\{\underbrace{v,\ldots,v}_{w}, \underbrace{1,\ldots,1}_{p-w}\},
\end{align*}
where the first $w$ diagonal entries represent low dimensional signal in the data, of magnitude $v>1$. A subset of the simulation results are presented in Table~\ref{table:K1results}, with complete results provided in Supplementary Table~S1.

For \texttt{pvclust} AU and BP values, summaries are reported for tests at the second and third nodes from the root, i.e. $j=(N-2)$ and $j=(N-3)$. For both SHC1 and SHC2, summaries are reported for the $p$-value at the root node of each simulated dataset. Under each set of simulation parameters, for each method, we report the number of replications with false positive calls using a significance threshold of $0.05$, as well as the mean $p$-value, and the median computing time of a single replication. For \texttt{pvclust}, a false positive was recorded if either of the two nodes was significant, and the mean $p$-value was calculated using both nodes. For a fair comparison of the computational times required by \texttt{pvclust} using 1000 bootstraps and the SHC procedures using 100 Monte Carlo simulations, we report the computational times of \texttt{pvclust} after scaling by $1/10$. Only a single computing time is reported for \texttt{pvclust}, as the implementation computes both AU and BP values simultaneously.

Since the data were generated from a single Gaussian distribution, we expect the SHC2 $p$-value at the root node to be approximately uniformly distributed over $[0, 1]$. In Table~\ref{table:K1results}, all methods show generally conservative behavior, making less false positive calls than expected by chance. The \texttt{pvclust} BP value (\texttt{pv}BP) shows the most strongly conservative behavior, reporting mean $p$-values close to $1$ for most settings. The remaining approaches, including the \texttt{pvclust} AU value (\texttt{pv}AU), and both SHC1 and SHC2, are consistently conservative across all settings considered. The conservative behavior of the classical SigClust procedure was previously described in \citet{Liu2008} and \citet{Huang2013} as being a result of the challenge of estimating the null eigenvalues and the corresponding covariance structure in the HDLSS setting \cite{Baik2006}. As both SHC1 and SHC2 rely on the same null covariance estimation procedure, this may also explain the generally conservative behavior observed in our proposed approaches. Both SHC approaches required substantially less computational time than \texttt{pvclust}, even after correcting for the larger number of bootstrap samples required by the method.

\subsection{Three Cluster Setting ($K=3$)}
\label{sim:K3}

\begin{table}[t!]\footnotesize
	\centering
	\begin{tabular}{rrl rrrrr rrrr}
\toprule
\multicolumn{3}{c}{parameters}&
	\multicolumn{5}{c}{$|\hat{K}=3|$}&
	\multicolumn{4}{c}{median time (sec.)}
\tabularnewline
\cmidrule(r){1-3} \cmidrule(r){4-8} \cmidrule(r){9-12}
	\multicolumn{1}{c}{$p$}&
	\multicolumn{1}{c}{$\delta$}&
	\multicolumn{1}{c}{arr.}&
	\multicolumn{1}{c}{\texttt{pv}AU}&
	\multicolumn{1}{c}{\texttt{pv}BP}&
	\multicolumn{1}{c}{SHC1}&
	\multicolumn{1}{c}{$\text{SHC2}_L$}&
	\multicolumn{1}{c}{$\text{SHC2}_2$}&
	\multicolumn{1}{c}{\texttt{pv}}&
	\multicolumn{1}{c}{SHC1}&
	\multicolumn{1}{c}{$\text{SHC2}_L$}&
	\multicolumn{1}{c}{$\text{SHC2}_2$}
\tabularnewline
\midrule
$   2$ & $ 3$ &    $\cdots$ & $-$  & $-$  &  18  &  0   &  29  & $-$       &  $ 2.42$ & $ 1.08$  &  $ 1.95$ \tabularnewline
$   2$ & $ 4$ &    $\cdots$ & $-$  & $-$  &  84  &  6   &  87  & $-$       &  $ 2.39$ & $ 1.08$  &  $ 1.92$ \tabularnewline
$1000$ & $ 8$ &    $\cdots$ & 0    & 0    &  0   &  5   &  66  & $231.62^*$ &  $79.85$ & $48.87$  &  $59.33$ \tabularnewline
$1000$ & $12$ &    $\cdots$ & 0    & 0    &  16  &  93  &  100 & $231.53^*$ &  $79.35$ & $49.14$  &  $59.21$ \tabularnewline
$1000$ & $20$ &    $\cdots$ & 13   & 0    &  70  &  79  &  99  & $231.55^*$ &  $78.62$ & $48.67$  &  $58.71$ \tabularnewline
\addlinespace
$   2$ & $ 4$ & $\triangle$ & $-$  & $-$  &  26  &  32  &  84  & $-$       &  $ 2.40$ & $ 1.07$  &  $ 1.97$ \tabularnewline
$   2$ & $ 5$ & $\triangle$ & $-$  & $-$  &  96  &  93  &  99  & $-$       &  $ 2.40$ & $ 1.06$  &  $ 1.94$ \tabularnewline
$1000$ & $ 8$ & $\triangle$ & 0    & 0    &  0   &  4   &  84  & $231.84^*$ &  $79.76$ & $49.19$  &  $59.21$ \tabularnewline
$1000$ & $12$ & $\triangle$ & 0    & 0    &  100 &  100 &  100 & $231.75^*$ &  $80.06$ & $49.43$  &  $59.17$ \tabularnewline
$1000$ & $20$ & $\triangle$ & 52   & 0    &  100 &  100 &  100 & $232.54^*$ &  $80.71$ & $49.29$  &  $59.58$ \tabularnewline
\bottomrule
\end{tabular}

	\caption{Simulation~\ref{sim:K3} ($K=3$). Number of replications identifying the correct number of significant clusters, median computation time over 100 replications. $*$: scaled by $1/10$.}
	\label{table:K3results}
\end{table}

We next consider the alternative setting in which datasets were drawn equally from three spherical Gaussian distributions each with covariance matrix $\textbf{I}_p$. The setting illustrates the simplest case for which significance must be attained at multiple nodes to discern the true clustering structure from the dendrogram using SHC. Two arrangements of the three Gaussian components were studied. In the first, the Gaussian components were placed along a line with distance $\delta$ between the means of neighboring components. In the second, the Gaussian components were placed at the corners of an equilateral triangle with side length $\delta$. Several values of $\delta$ were used to evaluate the relative power of each method across varying levels of signal. Both low ($p=2$) and high ($p=1000$) dimensional settings were also considered. For each dataset, 50 samples were drawn from each of the three Gaussian components. As in Simulation~\ref{sim:K1}, to make timing results comparable between \texttt{pvclust} and SHC, \texttt{pvclust} times are reported after scaling by $1/10$. Select simulation results are presented in Table~\ref{table:K3results}, with complete results presented in Supplementary Tables~S3 and~S4. We report the number of replications out of 100 for which each method detected statistically significant evidence of three ($\hat{K}=3$) clusters, as well as the median computation time across replications. For the two \texttt{pvclust} approaches, the numbers of predicted clusters were determined by the number of significant subtrees with at least $\ceil{(3/4)\cdot 50} = 38$ observations. This criterion was used to minimize the effect of small spurious clusters reported as being significant by the methods. For SHC1 and SHC2, the numbers of predicted clusters were determined by the resulting number of subtrees after cutting a dendrogram at all significant nodes identified while controlling the FWER at $0.05$.

In both arrangements of the components, the \texttt{pvclust} based methods showed substantially lower power than the proposed three approaches, with \texttt{pv}BP achieving no power at all. Across all settings reported in Table~\ref{table:K3results}, $\text{SHC2}_2$ consistently achieves the greatest power. The relative performance of $\text{SHC2}_L$ and SHC1 appears to depend on both the arrangement of the cluster components and the dimension of the dataset. When the components are arranged along a line, $\text{SHC2}_L$ outperforms SHC1 in the high dimensional setting, while the performance is reversed in the low dimensional setting. In contrast, when the components are placed in the triangular arrangement, $\text{SHC2}_L$ shows a slight advantage in both high and low dimensional settings. Timing results were comparable to those observed in Simulation~\ref{sim:K1}, with \texttt{pvclust} requiring substantially more time than the other approaches, even after scaling. Again, $\text{SHC2}_L$ required the least amount of computational time.

\subsection{Four Cluster Setting ($K=4$)}
\label{sim:K4}

\begin{table}[t!]\footnotesize
	\centering
	\begin{tabular}{rrl rrrrr rrrr}
\toprule
\multicolumn{3}{c}{parameters}&
	\multicolumn{5}{c}{$|\hat{K}=4|$}&
	\multicolumn{4}{c}{median time (sec.)}
\tabularnewline
\cmidrule(r){1-3} \cmidrule(r){4-8} \cmidrule(r){9-12}
	\multicolumn{1}{c}{$p$}&
	\multicolumn{1}{c}{$\delta$}&
	\multicolumn{1}{c}{arr.}&
	\multicolumn{1}{c}{\texttt{pv}AU}&
	\multicolumn{1}{c}{\texttt{pv}BP}&
	\multicolumn{1}{c}{SHC1}&
	\multicolumn{1}{c}{$\text{SHC2}_L$}&
	\multicolumn{1}{c}{$\text{SHC2}_2$}&
	\multicolumn{1}{c}{\texttt{pv}}&
	\multicolumn{1}{c}{SHC1}&
	\multicolumn{1}{c}{$\text{SHC2}_L$}&
	\multicolumn{1}{c}{$\text{SHC2}_2$}
\tabularnewline
\midrule
$   2$&$ 3$&square&$-$&$-$&$  3$&$  0$&$ 17$&$-$&$  2.54$&$ 1.16$&$ 2.06$\tabularnewline
$   2$&$ 4$&square&$-$&$-$&$ 78$&$ 12$&$ 90$&$-$&$  2.57$&$ 1.16$&$ 2.07$\tabularnewline
$1000$ & $ 8$ & square &      $ 0$ & $ 0$ & $  0$ & $  0$ & $ 75$ & $401.83^*$ & $110.50$ & $69.01$ & $82.59$ \tabularnewline
$1000$ & $10$ & square &      $ 0$ & $ 0$ & $ 97$ & $100$ & $100$ & $400.36^*$ & $110.79$ & $69.30$ & $82.67$ \tabularnewline
\addlinespace
$   3$&$ 4$&tetra.&$-$&$-$&$  0$&$  9$&$ 33$&$-$&$  2.84$&$ 1.28$&$ 2.25$\tabularnewline
$   3$&$ 5$&tetra.&$-$&$-$&$ 24$&$ 86$&$ 99$&$-$&$  2.40$&$ 1.09$&$ 2.03$\tabularnewline
$1000$ & $ 8$ & tetra. & $ 0$ & $ 0$ & $  0$ & $  0$ & $ 31$ & $402.00^*$ & $113.49$ & $71.03$ & $85.27$ \tabularnewline
$1000$ & $10$ & tetra. & $ 0$ & $ 0$ & $ 50$ & $ 99$ & $100$ & $399.85^*$ & $113.67$ & $71.17$ & $85.19$ \tabularnewline
\bottomrule
\end{tabular}
	\caption{Simulation~\ref{sim:K4} ($K=4$). Number of replications identifying the correct number of significant clusters, median computation time over 100 replications. $*$: scaled by $1/10$.}
	\label{table:K4results}
\end{table}

Finally, we consider the alternative setting in which datasets were drawn equally from four spherical Gaussian distributions each with covariance matrix $\textbf{I}_p$. Two arrangements of the Gaussian components were studied. In the first, the four components were placed at the vertices of a square with side length $\delta$. In the second, the four components were placed at the vertices of a regular tetrahedron, again with side length $\delta$. As in Simulation~\ref{sim:K3}, for each dataset, 50 samples were drawn from each of the Gaussian components. A representative subset of simulation results are presented in Table~\ref{table:K4results} for several values of $p$ and $\delta$, with complete results presented in Supplementary Tables~S5 and~S6. In the Supplementary Materials, we also include simulation results for a rectangular arrangement with side lengths $\delta$ and $(3/2)\cdot\delta$ (Supplementary Table~S7), and a stretched tetrahedral arrangement, also having side lengths $\delta$ and $(3/2)\cdot\delta$ (Supplementary Table~S8).

The results presented in Table~\ref{table:K4results} largely support the results observed in Simulation~\ref{sim:K3}. Again, the \texttt{pv}AU and \texttt{pv}BP values provide little power to detect significant clustering in the data, while $\text{SHC2}_2$ consistently achieves the greatest power. Additionally, the relative performance of $\text{SHC2}_L$ and SHC1 again depends on the arrangement of the components and the dimension of the dataset. In the square arrangement, while SHC1 performs better in the low dimensional setting, the approaches perform equally well in high dimensions. However, in the tetrahedral arrangement, $\text{SHC2}_L$ achieves substantially greater power than SHC1 in both high and low dimensional settings.

%

\section{REAL DATA ANALYSIS}
\label{realdata}
To further demonstrate the power of SHC, we apply the approach to two cancer gene expression datasets. We first consider a dataset of 300 tumor samples drawn from three distinct cancer types: head and neck squamous cell carcinoma (HNSC), lung squamous cell carcinoma (LUSC), and lung adenocarcinoma (LUAD). As distinct cancers, we expect observations from the three groups to be easily separated by hierarchical clustering and detected by SHC. In the second dataset, we consider a cohort of 337 breast cancer (BRCA) samples, previously categorized into five molecular subtypes \cite{Parker2009}. The greater number of subpopulations, as well as the more subtle differences between them, makes this dataset more challenging than the first. In both examples, the data were clustered using Ward's linkage and the $\text{SHC2}_2$ approach was implemented as described in Section~\ref{simulations} using 1000 simulations at each node. The FWER controlling procedure of Section~\ref{methodology:FWER} was applied with $\alpha=0.05$.

\subsection{Multi-Cancer Gene Expression Dataset}
\begin{figure}[!t]
	\centering
	\includegraphics[width=\textwidth]{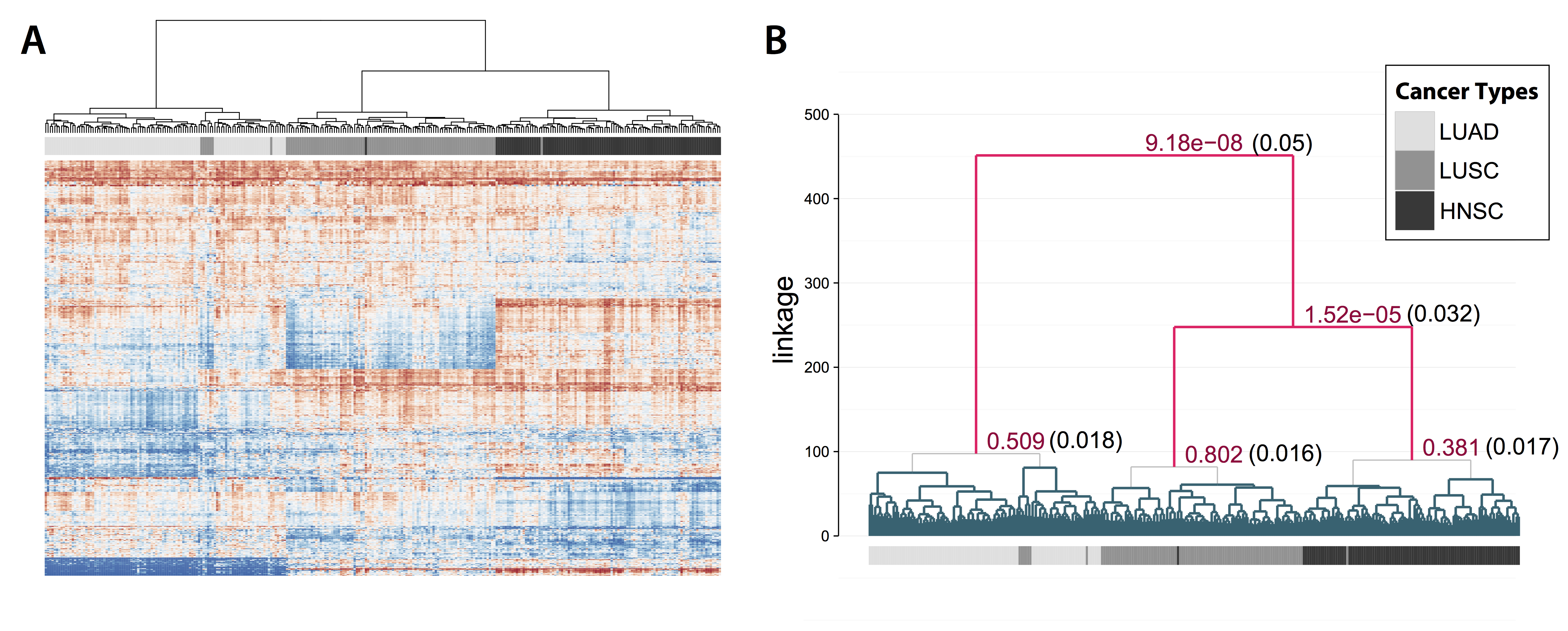}%
	\caption[Example of hierarchical clustering and a dendrogram.]{Analysis of gene expression for 300 LUAD, LUSC, and HNSC samples. (A) Heatmap of log-transformed gene expression for the 300 samples (columns), clustered by Ward's linkage. (B) Dendrogram with corresponding SHC $p$-values (red) and $\alpha^*$ cutoffs (black) given only at nodes tested according to the FWER controlling procedure at $\alpha=0.05$.}
	\label{fig:multicancer}
\end{figure}

A dataset of 300 samples was constructed by combining 100 samples from each of HNSC, LUSC and LUAD, all obtained from The Cancer Genome Atlas (TCGA) project \cite{Cancer2012,Cancer2014}. Gene expression was estimated for 20,531 genes from RNA-seq data using RSEM \cite{Li2011a}, as described in the TCGA RNA-seq v2 pipeline (\url{https://wiki.nci.nih.gov/display/TCGA/RNASeq+Version+2}). To adjust for technical effects of the data collection process, expression values were first normalized using the upper-quartile procedure of \citet{Bullard2010}. Then, all expression values of zero were replaced by the smallest non-zero expression value across all genes and samples. A subset of 500 most variably expressed genes were selected according to the median absolute deviation about the median (MAD) expression across all samples. Finally, SHC was applied to the log-transformed expression levels at the 500 most variable loci. Similar results were also obtained when using the 100, 1000, and 2000 most variable genes. 

In Figure~\ref{fig:multicancer}A, the log-transformed expression values are visualized using a heatmap, with rows corresponding to genes, and columns corresponding to samples. Lower and higher expression values are shown in blue and red, respectively. For easier visual interpretation, rows and columns of the heatmap were independently clustered using Ward's linkage clustering. The corresponding dendrogram and cancer type labels are shown above the heatmap. The dendrogram and labels in Panel A of Figure~\ref{fig:multicancer} are reproduced in Panel B, along with the SHC $p$-values (red) and modified significance cutoffs (black) at nodes tested according to the FWER controlling procedure. Branches corresponding to statistical significant nodes and untested nodes are shown in red and blue, respectively. Ward's linkage clustering correctly separates the three cancer types, with the exception of seven LUSC samples clustered with the LUAD samples, one LUSC sample clustered with HNSC, and one HNSC sample clustered with LUSC. Interestingly, the LUSC and HNSC samples cluster prior to joining with the LUAD samples, suggesting the greater molecular similarity between squamous cell tumors of different sites, than different cancers of the lung. This agrees with the recently identified genomic similarity of the two tumors reported in \citet{Hoadley2014}. Furthermore, we note that no HNSC and LUAD samples are jointly clustered, highlighting the clear difference between tumors of both distinct histology and site. As shown in Figure~\ref{fig:multicancer}B, statistically significant evidence of clustering was determined at the top two nodes, with respective Gaussian-fit $p$-values $9.18e-8$ and $1.52e-5$ at the modified significance cutoffs, $\alpha^*_{299}=0.05$ and $\alpha^*_{298}=0.032$. Additionally, the three candidate nodes corresponding to splitting each of the cancer types all give insignificant results, suggesting no further clustering in the cohort. Finally, we note that when analyzed using \texttt{pvclust}, no statistically significant evidence of clustering was found, with AU $p$-values of $0.26$, $0.28$, and $0.13$ obtained at the three nodes corresponding to primarily LUAD, LUSC and HNSC samples.

\subsection{BRCA Gene Expression Dataset}
As a second example, we consider a microarray gene expression dataset from 337 BRCA samples. The dataset was compiled, filtered and normalized as described in \citet{Prat2010} and obtained from the University of North Carolina (UNC) Microarray Database (\url{https://genome.unc.edu/pubsup/clow/}). Gene expression was analyzed for a subset of 1645 well-chosen \emph{intrinsic} genes identified in \citet{Prat2010}. We evaluate the ability of our approach to detect biologically relevant clustering based on five molecular subtypes: luminal A (LumA), luminal B (LumB), basal-like, normal breast-like, and HER2-enriched \cite{Parker2009}. The dataset is comprised of 97 LumA, 54 LumB, 91 basal-like, 47 normal breast-like, and 48 HER2-enriched samples. 

\begin{figure}[!t]
	\centering
	\includegraphics[width=.9\textwidth]{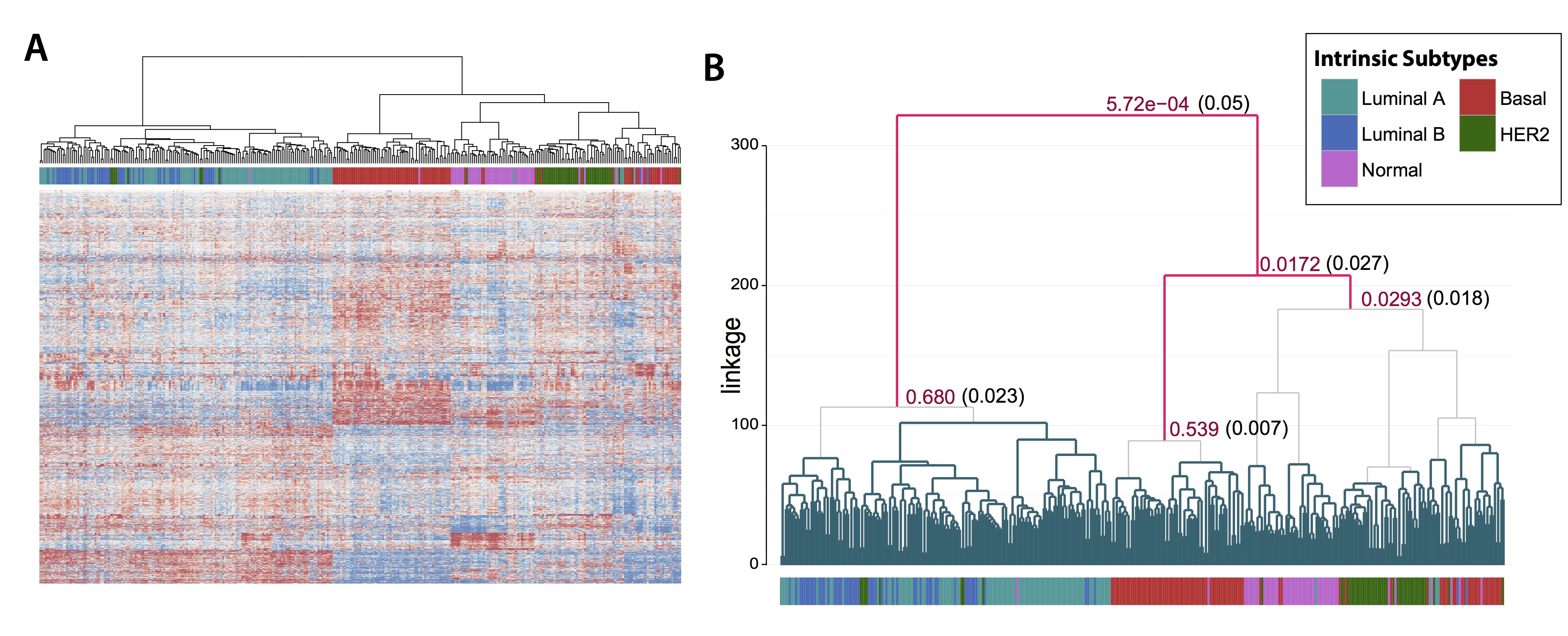}
	\caption[Example of hierarchical clustering and a dendrogram.]{Analysis of gene expression for 337 BRCA samples. (A) Heatmap of gene expression for the 337 samples (columns) clustered by Ward's linkage. (B) Dendrogram with corresponding SHC $p$-values (red) and $\alpha^*$ cutoffs (black) given only at nodes tested according to the FWER controlling procedure at $\alpha=0.05$.}
	\label{fig:BRCA}
\end{figure}

The expression dataset is shown as a heatmap in Figure~\ref{fig:BRCA}A, with the corresponding dendrogram and subtype labels reproduced in Figure~\ref{fig:BRCA}B. The corresponding SHC $p$-values (red) and modified significance thresholds (black) are again given at only nodes tested while controlling the FWER at $\alpha=0.05$. SHC identifies at least three significantly differentiated clusters in the dataset, primarily corresponding to luminal (LumA and LumB), basal-like, and all remaining subtypes. At the root node, the LumA and LumB samples are separated from the remaining subtypes with a $p$-value of $5.72e-4$ at a threshold of $\alpha^*_{336}=0.05$. However, Ward's linkage clustering and SHC are unable to identify significant evidence of clustering between the two luminal subtypes. The difficultly of clustering LumA and LumB subtypes based on gene expression was previously described in \citet{Mackay2011}. Next, the majority of basal-like samples are separated from the remaining set of observations, with a $p$-value of $0.0172$ at a cutoff of $\alpha^*_{335}=0.027$. The remaining HER2-enriched, normal breast-like and basal-likes samples show moderate separation by Ward's linkage clustering. However, controlling the FWER at $\alpha=0.05$, the subsequent node is non-significant, with a $p$-value of $0.0293$ against a corrected threshold of $\alpha^*_j=0.0180$. This highlights the difficulty of assessing statistical significance in the presence of larger numbers of clusters, while controlling for multiple testing.

\section{DISCUSSION}
\label{discussion}
While hierarchical clustering has become widely popular in practice, few methods have been proposed for assessing the statistical significance of a hierarchical partition. SHC was developed to address this problem, using a sequential testing and FWER controlling procedure. Through an extensive simulation study, we have shown that SHC provides competitive results compared to existing methods. Furthermore, in applications to two gene expression datasets, we showed that the approach is capable of identifying biologically meaningful clustering. 

In this paper, we focused on the theoretical and empirical properties of SHC using Ward's linkage. However, there exist several different approaches to hierarchical clustering, and Ward's linkage may not always be the most appropriate choice. In these situations, as mentioned in Section~\ref{methodology}, SHC may be implemented with other linkage and dissimilarity functions which satisfy mean shift and rotation invariance. Further investigation is necessary to fully characterize the behavior of the approach for different hierarchical clustering procedures.

Some popular choices of dissimilarity, such as those based on Pearson correlation of the covariates between pairs of samples, fail to satisfy the necessary mean shift and rotation invariance properties in the original covariate space. As a consequence, the covariance of the Gaussian null distribution must be fully estimated, and cannot be approximated using only the eigenvalues of the sample covariance matrix. When $N\gg p$, the SHC method can still be applied by estimating the complete covariance matrix. However, in HDLSS settings, estimation of the complete covariance matrix can be difficult and computationally expensive. A possible direction of future work is the development of a computationally efficient procedure for non-invariant hierarchical clustering procedures.

\section*{APPENDIX: PROOFS}
\label{proofs}

\subsection*{Proof of Theorem~\ref{thm:wardsplit}}
Let $d_W(\cdot,\cdot)$ denote the Ward's linkage function defined over sets of observation indices. Additionally, let $\bbX^{(1)}$ and $\bbX^{(2)}$ denote $n$ and $m$ samples drawn from two Gaussian components with distributions $N(\boldmu_1, \sigma_1^2 \bold{I}_p)$ and $N(\boldmu_2, \sigma_2^2 \bold{I}_p)$, with corresponding observation index sets, $\mathbb{C}^{(1)}$ and $\mathbb{C}^{(2)}$. For $k=1,2$, let $C^{(k)}_0$, $C^{(k)}_1$ and $C^{(k)}_2$ denote subsets of $\mathbb{C}^{(k)}$, where $C^{(k)}_1$ and $C^{(k)}_2$ are necessarily disjoint. Let $n_0 = |C^{(1)}_0|$, $n_1 = |C^{(1)}_1|$, $n_2 = |C^{(1)}_2|$, $m_0 = |C^{(2)}_0|$, $m_1 = |C^{(2)}_1|$ and $m_2 = |C^{(2)}_2|$ denote the size of each subset. Finally, let $\bX^{(k)}_0$, $\bX^{(k)}_1$, and $\bX^{(k)}_2$, denote the corresponding subsets of $\bbX^{(k)}$, with corresponding sample means, $\overline\bX^{(k)}_0$, $\overline\bX^{(k)}_1$, and $\overline\bX^{(k)}_2$.

Consider the two events: $A = \{ \max d_W (C^{(1)}_1, C^{(1)}_2) < \min d_W (C^{(1)}, C^{(2)}) \}$, and $B = \{ \max d_W (C^{(2)}_1, C^{(2)}_2) < \min d_W (C^{(1)}, C^{(2)}) \}$, where maxima and minima are taken with respect to the possible values of $C^{(k)}_1$, $C^{(k)}_2$, and $C^{(k)}$. Note that the joint occurrence of $A$ and $B$ is sufficient for correctly separating observations from the two components at the root node. It therefore suffices to show that $P(A \cap B) \rightarrow 1$ as $p \rightarrow \infty$. 

For some $0 < a_1 \leq a_2$, define the following events: $E_{1} = \{ \max d_W (C^{(1)}_1, C^{(1)}_2) < a_1\cdot p \}$, $E_{2} = \{ \max d_W (C^{(2)}_1, C^{(2)}_2) < a_1\cdot p \}$, $E_{3} = \{ \min d_W (C^{(1)}_0, C^{(2)}_0) > a_2\cdot p \}$. Note that the probability of the joint event $(A \cap B)$ can be bounded below by:
\begin{align*}
	P(A \cap B)
		&= 1 - P(A^C \cup B^C) \\
		&\geq 1 - \big( P(E_1^C) + P(E_2^C) + P(E_3^C) \big).
\end{align*}
We complete the proof by showing that $P(E_1^C)$, $P(E_2^C)$, $P(E_3^C)$ all tend to 0 as $p \rightarrow \infty$.

By Lemmas 3 and 4 of \citet{Borysov2014} for $a_1>2\sigma_1^2$, we have
\begin{align*}
	P(E_1^C) &= P(\max d_W (C^{(1)}_1, C^{(1)}_2)  > a_1 \cdot p) \\
		&= P\left(\max \frac{2 n_1 n_2}{n_1+n_2} %
				\left\| \overline\bX^{(1)}_1 - \overline\bX^{(1)}_2 \right\|^2 > a_1 \cdot p %
			\right) \\
		&\leq 3^n P\left(\left\|\left(\frac{2 n_1 n_2}{n_1+n_2}\right)^{1/2} %
				\left(\overline\bX^{(1)}_1 - \overline\bX^{(1)}_2\right)\right\|^2 > a_1 \cdot p %
			\right) \\
		&\leq e^{-c_1 p + n\log 3},
\end{align*}
where $c_1 = a_1 / \sigma_1^2 - (1+ \log(a_1/\sigma_1^2))$. Note that since $c_1>0$ and $n = o(p)+p^\alpha$ for some $\alpha\in(0,1)$, $P(E_1^C)\rightarrow0$ as $p\rightarrow\infty$. Similarly, for $a_2>2\sigma_2^2$, we have
\begin{align*}
	P(E_2^C) &\leq e^{-c_2 p + m\log 3},
\end{align*}
where $c_2 = a_2 / \sigma_2^2 - (1+ \log(a_2/\sigma_2^2))$, such that $P(E_2^C)\rightarrow0$ as $p\rightarrow\infty$. Finally, to bound $P(E_3^C)$, we make use of Lemmas 2 and 4 of \citet{Borysov2014}:
\begin{align*}
	P(E^C_3) &= P(\min d_W (C^{(1)}_0, C^{(2)}_0) < a_2 \cdot p) \\
		&\leq \sum_{i=1}^{n} \sum_{j=1}^{m} P \left(\frac{2ij}{i+j} %
			\left\|\overline\bX^{(1)}_0 - \overline\bX^{(2)}_0\right\|^2 < a_2 \cdot p\right) \\
		&\leq 2^{n+m} \max_{i\leq n,\ j\leq m}  P \left( \frac{2ij}{i+j} %
			\left\|\overline\bX^{(1)}_0 - \overline\bX^{(2)}_0\right\|^2 < a_2 \cdot p\right).
\end{align*}
Suppose that $i$ and $j$ are fixed, and let $\mu^2 = p^{-1}(\tfrac{2ij}{i+j})\|\boldmu_1 - \boldmu_2\|^2$, $\mu_k = (\tfrac{2ij}{i+j})^{1/2}(\boldmu_{1,k} - \boldmu_{2,k})$, $\sigma^2 = (\tfrac{2ij}{i+j})(\tfrac{i\sigma_2^2 + j\sigma_1^2}{ij})$. Then, using the result of Lemma 2 of \citet{Borysov2014}, for $0<a_2<\sigma^2+\mu^2$, we have
\begin{align*}
	P \left( \frac{2ij}{i+j} \left\|\overline\bX^{(1)}_0 - \overline\bX^{(2)}_0\right\|^2 < a_2 \cdot p\right)
		&\leq e^{-c_3 p},
\end{align*}
where $c_3 = {(a_2 - \sigma^2 - \mu^2)^2} / %
		{(6\sigma^4 + 12 \sigma^2\mu^2 + 2p^{-1}\sum_{k=1}^p \mu_k^4)}$.

Using the fourth moment bound of~(\ref{eq:L4bound}), and the fact that $n=o(p)+p^\alpha$, $m=o(p)+p^\beta$, we have that $P(E_3^C)\rightarrow0$ as $p\rightarrow\infty$. Thus, for $a_1 > 2\sigma_1^2$, $P(E^C_1) \rightarrow 0$, for $a_1 > 2\sigma_2^2$, $P(E^C_2) \rightarrow 0$, and for $a_2 < (\tfrac{2nm}{n+m})(\tfrac{\sigma_1^2}{n} + \tfrac{\sigma_2^2}{m} + \tfrac{\|\boldmu_1 - \boldmu_2\|^2}{p})$, $P(E^C_3) \rightarrow 0$. Combining the necessary inequalities on $a_1, a_2$, we obtain the stated condition:
\begin{align*}
	2\sigma^2_2 
		< a_1 &\leq a_2 \\
		&< \frac{2nm}{n+m} \left( \frac{\sigma^2_1}{n} + \frac{\sigma^2_2}{m} + \frac{\|\boldmu_1 - \boldmu_2\|^2}{p} \right) \\
	\frac{1}{n}(\sigma^2_2 - \sigma^2_1)
		&< \frac{\|\boldmu_1 - \boldmu_2\|^2}{p}.
\end{align*}
%

\subsection*{Proof of Theorem~\ref{thm:wardpower}}
Let $d_W(\cdot,\cdot)$ denote the Ward's linkage function. Further, let $\bbX^{(1)}$ and $\bbX^{(2)}$ denote the $n$ and $m$ observations from the first and second Gaussian components with distributions $N(\boldmu_1, \sigma_1^2 \bold{I}_p)$ and $N(\boldmu_2, \sigma_2^2 \bold{I}_p)$. Assume that $\boldmu_1$, $\boldmu_2$, $\sigma_1^2$ and $\sigma_2^2$ are known. Then, the theoretical best fit Gaussian to the mixture distribution is equivalent (up to a mean shift and rotation) to $N(\bold{0}, \widehat\boldSigma)$, where
\begin{align*}
	\widehat\boldSigma &= \text{diag}\{\widehat\lambda_k\}_{k=1}^p \\
	\widehat\lambda_1 &= \frac{nm}{n+m}\left((n+m)^{-1} \|\boldmu_1 - \boldmu_2\|^2 + \frac{\sigma_1^2}{m} + \frac{\sigma_2^2}{n}\right) \\
	\widehat\lambda_k &= \frac{nm}{n+m}\left(\frac{\sigma_1^2}{m} + \frac{\sigma_2^2}{n}\right), \ \ \text{for } k \geq 2,
\end{align*}
where the $\widehat\lambda_k$ are derived by the formula for the variance of a univariate mixture of Gaussians. Let $\bbX^{(3)}$ denote a sample of $n+m$ observations drawn from $N(\bold{0}, \widehat\boldSigma)$. Let $\mathbb{C}^{(k)}$ denote the corresponding observation indices for $k=1,2,3$. Additionally, let $C^{(3)}_1$ and $C^{(3)}_2$ denote disjoint subsets of $\mathbb{C}^{(3)}$, and let $r_1 = |C^{(3)}_1|$ and $r_2 = |C^{(2)}_2|$ denote the size of the subsets. Finally, let $\bX^{(3)}_1$ and $\bX^{(3)}_2$ denote the corresponding subsets of $\bbX^{(3)}$ with means $\overline\bX^{(3)}_1$ and $\overline\bX^{(3)}_2$.

Consider the event: $D = \{ \max d_W (C^{(3)}_1, C^{(3)}_2) < d_W (\mathbb{C}^{(1)}, \mathbb{C}^{(2)}) \}$, where the maximum is taken with respect the possible values of $C^{(3)}_1$ and $C^{(3)}_2$. By Theorem~\ref{thm:wardsplit} we have that, asymptotically, Ward's linkage clustering achieves the correct partition of $\mathbb{C}^{(1)}$ and $\mathbb{C}^{(2)}$. Therefore, $D$ is precisely the event that a linkage value simulated from the null distribution is less than the observed linkage value. The proof is completed by showing $P(D) \rightarrow 1$ as $p \rightarrow \infty$. That is, we wish to show that the empirical $p$-value tends to $0$ as $p\rightarrow\infty$. 

For some $a>0$, define the following events: $E_{4} = \{ \max d_W (C^{(3)}_1, C^{(3)}_2) < a \cdot p \}$, and $E_{5} = \{ d_W (\mathbb{C}^{(1)}, \mathbb{C}^{(2)}) > a \cdot p \}$. Note that $P(D)$ can be bounded below by:
\begin{align*}
	P(D)
		&= 1 - P(D^C) \\
		&\geq 1 - \big( P(E_4^C) + P(E_5^C) \big).
\end{align*}
Thus, it suffices to show the probabilities of $E_4^C$, $E_5^C$,both tend to 0 as $p \rightarrow \infty$. 

First, we state a generalization of Lemma 3 from \citet{Borysov2014} for Gaussian distributions with diagonal covariance. 
\begin{lemma}\label{lemma:genLem3}
Suppose $n$ independent observations, $\bbX$, are drawn from the $p$-dimensional Gaussian distribution, $N(\boldmu, \boldSigma)$, where $\boldSigma$ is a diagonal matrix with diagonal entries $\{\lambda_k\}_{k=1}^p$. Define scalars $\mu^2 = p^{-1}\|\boldmu\|^2$, $\overline\lambda = p^{-1}\sum_{k=1}^p \lambda_k$, and let $a> \overline\lambda + \mu^2$. Then, for any $0<i\leq n$,
\begin{align*}
	P(\|X_i\|^2 > a\cdot p) \leq e^{-cp},
\end{align*}
where $c = \left[a + \mu^2 - \sqrt{\overline\lambda^2 + 4\mu^2a} + \overline\lambda\left( \frac{\overline\lambda + \sqrt{\overline\lambda^2 + 4\mu^2a}}{2a} \right) \right] / \overline\lambda$.
\end{lemma}
The proof of Lemma~\ref{lemma:genLem3} is omitted as it follows exactly as that of Lemma~3 from \citet{Borysov2014}. By Lemma~\ref{lemma:genLem3} given above and Lemma~4 of \citet{Borysov2014} , for $a > 2 \overline{\widehat\lambda}$, where $\overline{\widehat\lambda} = p^{-1} \sum_{k=1}^p \widehat\lambda_k$, we have
\begin{align*}
	P(E_4^C) 
		&= P(\max d_W (C^{(3)}_1, C^{(3)}_2)  > a \cdot p) \\
		&= P\left(\max \frac{2 r_1 r_2}{r_1+r_2} %
				\left\| \overline\bX^{(3)}_1 - \overline\bX^{(3)}_2 \right\|^2 > a \cdot p %
			\right) \\
		&\leq 3^{n+m} P\left(\left\|\left(\frac{2 r_1 r_2}{r_1+r_2}\right)^{1/2} %
				\left(\overline\bX^{(3)}_1 - \overline\bX^{(3)}_2\right)\right\|^2 > a \cdot p %
			\right) \\
		&\leq e^{-c_4 p + (n+m)\log 3},
\end{align*}
where $c_4 = a / \overline{\widehat\lambda} - (1+ \log(a/\overline{\widehat\lambda}))$. As for $P(E_1^C)$ from the proof of Theorem~\ref{thm:wardsplit}, we have that as $p\rightarrow\infty$, $P(E_4^C)\rightarrow 0$ as $p\rightarrow0$. Next, using an argument similar to the one presented above for $P(E_3^C)\rightarrow0$, we show that $P(E_5^C)\rightarrow0$. Let $\mu^2 = p^{-1}(\tfrac{2nm}{n+m})\|\boldmu_1 - \boldmu_2\|^2$, $\mu_k = (\tfrac{2nm}{n+m})^{1/2}(\boldmu_{1,k} - \boldmu_{2,k})$, $\sigma^2 = (\tfrac{2nm}{n+m})(\tfrac{i\sigma_2^2 + j\sigma_1^2}{nm})$. Then, by Lemmas 2 and 4 of \citet{Borysov2014}, for $0<a<\sigma^2 + \mu^2$, we have
\begin{align*}
	P(E^C_4) &= P(d_W (\mathbb{C}^{(1)}, \mathbb{C}^{(2)}) < a \cdot p) \\
		&= P \left(\frac{2nm}{n+m} %
			\left\|\overline\bX^{(1)} - \overline\bX^{(2)}\right\|^2 < a \cdot p\right) \\
		&\leq e^{-c_5 p},
\end{align*}
where $c_5 = {(a - \sigma^2 - \mu^2)^2} / %
		{(6\sigma^4 + 12 \sigma^2\mu^2 + 2p^{-1}\sum_{k=1}^p \mu_k^4)}$. As for $P(E_3^C)$ from the proof of Theorem~\ref{thm:wardsplit}, we have $P(E_5^C)\rightarrow0$ as $p\rightarrow\infty$. Thus, for $a>2\overline{\widehat\lambda}$, $P(E_4^C)\rightarrow0$, and for $a<(\tfrac{2nm}{n+m})(\tfrac{\sigma^2_1}{n} + \tfrac{\sigma^2_2}{m} + \tfrac{\|\boldmu_1 - \boldmu_2\|^2}{p})$, $P(E_5^C)\rightarrow0$. Combining the two inequalities on $a$, we obtain the stated condition:
\begin{align*}
	2p^{-1}\sum_{k=1}^p \widehat\lambda_k 
		&< a \\
		&< \left(\frac{2nm}{n+m}\right)%
			\left(\frac{\sigma^2_1}{n} + \frac{\sigma^2_2}{m} + \frac{\|\boldmu_1 - \boldmu_2\|^2}{p}\right) \\
	\frac{\sigma_1^2}{m} + \frac{\sigma_2^2}{n} + %
		(n+m)^{-1}\cdot\frac{\|\boldmu_1 - \boldmu_2\|^2}{p}
		&< \frac{\sigma^2_1}{n} + \frac{\sigma^2_2}{m} + \frac{\|\boldmu_1 - \boldmu_2\|^2}{p} \\
	\frac{(m^2-n^2)(\sigma_2^2 - \sigma_1^2)}{nm(n+m-1)}%
		&< \frac{\|\boldmu_1 - \boldmu_2\|^2}{p}.
\end{align*}
%


\bibliographystyle{apalike}
\bibliography{references}


\clearpage
\section*{SUPPLEMENTARY MATERIALS}
\label{supplement}
\beginsupplement

\onehalfspacing

\begin{table}[h!]\footnotesize
	\centering
	\begin{tabular}{ccc rrrrr rrrr}
\toprule
\multicolumn{3}{c}{parameters}&
	\multicolumn{5}{c}{$|p\text{-value}<0.05|$ (mean $p$-value)}&
	\multicolumn{4}{c}{median time (sec.)}
\tabularnewline
\cmidrule(r){1-3} \cmidrule(r){4-8} \cmidrule(r){9-12}
\multicolumn{1}{c}{$n$}&
	\multicolumn{1}{c}{$w$}&
	\multicolumn{1}{c}{$v$}&
	\multicolumn{1}{c}{\texttt{pv}AU}&
	\multicolumn{1}{c}{\texttt{pv}BP}&
	\multicolumn{1}{c}{SHC1}&
	\multicolumn{1}{c}{$\text{SHC2}_L$}&
	\multicolumn{1}{c}{$\text{SHC2}_2$}&
	\multicolumn{1}{c}{\texttt{pv}}&
	\multicolumn{1}{c}{SHC1}&
	\multicolumn{1}{c}{$\text{SHC2}_L$}&
	\multicolumn{1}{c}{$\text{SHC2}_2$}
\tabularnewline
\midrule
$ 50$  &  $1$ & $1$ &    $ 0\ (1.00)$  &  $0\ (1.00)$  &  $0\ (1.00)$  &  $0\ (1.00)$  &  $0\ (1.00)$  &  $ 25.17$  &  $17.48$  &  $ 9.62$  &  $11.98$ \tabularnewline
$ 50$  &  $1$ & $10$ &   $ 0\ (1.00)$  &  $0\ (1.00)$  &  $0\ (1.00)$  &  $0\ (1.00)$  &  $0\ (1.00)$  &  $ 25.69$  &  $16.75$  &  $ 9.58$  &  $11.91$ \tabularnewline
$ 50$  &  $1$ & $25$ &   $ 0\ (0.95)$  &  $0\ (1.00)$  &  $0\ (0.89)$  &  $0\ (0.91)$  &  $0\ (0.70)$  &  $ 24.16$  &  $15.87$  &  $ 9.40$  &  $11.32$ \tabularnewline
$ 50$  &  $1$ & $100$ &  $ 3\ (0.61)$  &  $0\ (0.98)$  &  $0\ (0.60)$  &  $0\ (0.49)$  &  $0\ (0.46)$  &  $ 24.59$  &  $17.24$  &  $ 9.96$  &  $12.22$ \tabularnewline
$ 50$  &  $1$ & $500$ &  $ 4\ (0.28)$  &  $0\ (0.85)$  &  $0\ (0.53)$  &  $0\ (0.47)$  &  $1\ (0.44)$  &  $ 24.46$  &  $17.53$  &  $ 9.93$  &  $12.03$ \tabularnewline
$ 50$  &  $1$ & $1000$ & $13\ (0.19)$  &  $0\ (0.74)$  &  $0\ (0.57)$  &  $0\ (0.50)$  &  $0\ (0.47)$  &  $ 23.40$  &  $16.97$  &  $ 9.81$  &  $11.42$ \tabularnewline
\addlinespace
$100$  &  $1$ & $1$ &    $ 0\ (1.00)$  &  $0\ (1.00)$  &  $0\ (1.00)$  &  $0\ (1.00)$  &  $0\ (1.00)$  &  $ 85.48$  &  $37.28$  &  $21.69$  &  $26.12$ \tabularnewline
$100$  &  $1$ & $10$ &   $ 0\ (1.00)$  &  $0\ (1.00)$  &  $0\ (1.00)$  &  $0\ (1.00)$  &  $0\ (0.98)$  &  $ 88.23$  &  $38.71$  &  $21.95$  &  $26.97$ \tabularnewline
$100$  &  $1$ & $25$ &   $ 0\ (1.00)$  &  $0\ (1.00)$  &  $0\ (0.88)$  &  $0\ (0.59)$  &  $0\ (0.46)$  &  $ 81.09$  &  $38.12$  &  $22.19$  &  $26.86$ \tabularnewline
$100$  &  $1$ & $100$ &  $ 0\ (0.95)$  &  $0\ (1.00)$  &  $0\ (0.65)$  &  $0\ (0.46)$  &  $0\ (0.45)$  &  $ 87.60$  &  $37.86$  &  $22.18$  &  $28.07$ \tabularnewline
$100$  &  $1$ & $500$ &  $ 4\ (0.50)$  &  $0\ (0.95)$  &  $0\ (0.63)$  &  $0\ (0.49)$  &  $0\ (0.47)$  &  $ 82.48$  &  $38.35$  &  $23.46$  &  $27.62$ \tabularnewline
$100$  &  $1$ & $1000$ & $ 5\ (0.37)$  &  $0\ (0.91)$  &  $1\ (0.62)$  &  $0\ (0.49)$  &  $1\ (0.46)$  &  $ 84.30$  &  $39.03$  &  $22.65$  &  $27.49$ \tabularnewline
\addlinespace
$200$  &  $1$ & $1$ &    $ 0\ (1.00)$  &  $0\ (1.00)$  &  $0\ (1.00)$  &  $0\ (1.00)$  &  $0\ (1.00)$  &  $318.44$  &  $83.78$  &  $51.00$  &  $63.95$ \tabularnewline
$200$  &  $1$ & $10$ &   $ 0\ (1.00)$  &  $0\ (1.00)$  &  $0\ (1.00)$  &  $0\ (0.86)$  &  $1\ (0.68)$  &  $313.89$  &  $86.42$  &  $51.24$  &  $63.27$ \tabularnewline
$200$  &  $1$ & $25$ &   $ 0\ (1.00)$  &  $0\ (1.00)$  &  $0\ (0.97)$  &  $0\ (0.50)$  &  $0\ (0.49)$  &  $300.94$  &  $86.42$  &  $50.56$  &  $63.38$ \tabularnewline
$200$  &  $1$ & $100$ &  $ 0\ (1.00)$  &  $0\ (1.00)$  &  $0\ (0.77)$  &  $0\ (0.49)$  &  $0\ (0.48)$  &  $290.64$  &  $81.59$  &  $48.98$  &  $59.55$ \tabularnewline
$200$  &  $1$ & $500$ &  $ 0\ (0.93)$  &  $0\ (1.00)$  &  $0\ (0.69)$  &  $0\ (0.49)$  &  $0\ (0.46)$  &  $270.83$  &  $80.69$  &  $47.29$  &  $59.16$ \tabularnewline
$200$  &  $1$ & $1000$ & $ 0\ (0.76)$  &  $0\ (0.98)$  &  $0\ (0.72)$  &  $0\ (0.51)$  &  $1\ (0.49)$  &  $264.99$  &  $80.24$  &  $46.28$  &  $58.00$ \tabularnewline
\addlinespace
$ 50$  &  $5$ & $10$ &   $ 0\ (1.00)$  &  $0\ (1.00)$  &  $0\ (1.00)$  &  $0\ (1.00)$  &  $0\ (1.00)$  &  $ 21.40$  &  $14.53$  &  $ 8.24$  &  $ 9.72$ \tabularnewline
$ 50$  &  $5$ & $25$ &   $ 0\ (0.97)$  &  $0\ (1.00)$  &  $0\ (0.92)$  &  $0\ (0.92)$  &  $0\ (0.76)$  &  $ 21.45$  &  $14.66$  &  $ 8.29$  &  $ 9.80$ \tabularnewline
$ 50$  &  $5$ & $100$ &  $ 4\ (0.85)$  &  $0\ (1.00)$  &  $0\ (0.84)$  &  $0\ (0.73)$  &  $0\ (0.68)$  &  $ 21.25$  &  $14.48$  &  $ 8.27$  &  $ 9.91$ \tabularnewline
$ 50$  &  $5$ & $500$ &  $ 7\ (0.66)$  &  $0\ (0.99)$  &  $0\ (0.91)$  &  $0\ (0.77)$  &  $0\ (0.79)$  &  $ 21.30$  &  $14.53$  &  $ 8.28$  &  $ 9.82$ \tabularnewline
$ 50$  &  $5$ & $1000$ & $ 6\ (0.58)$  &  $0\ (0.99)$  &  $0\ (0.92)$  &  $0\ (0.75)$  &  $0\ (0.79)$  &  $ 21.31$  &  $14.60$  &  $ 8.40$  &  $ 9.95$ \tabularnewline
\addlinespace
$100$  &  $5$ & $10$ &   $ 0\ (1.00)$  &  $0\ (1.00)$  &  $0\ (1.00)$  &  $0\ (1.00)$  &  $0\ (0.92)$  &  $ 73.15$  &  $33.96$  &  $18.93$  &  $23.42$ \tabularnewline
$100$  &  $5$ & $25$ &   $ 0\ (1.00)$  &  $0\ (1.00)$  &  $0\ (0.96)$  &  $0\ (0.73)$  &  $1\ (0.67)$  &  $ 72.81$  &  $34.26$  &  $19.04$  &  $23.45$ \tabularnewline
$100$  &  $5$ & $100$ &  $ 0\ (1.00)$  &  $0\ (1.00)$  &  $0\ (0.95)$  &  $0\ (0.72)$  &  $0\ (0.72)$  &  $ 72.71$  &  $34.30$  &  $19.06$  &  $23.54$ \tabularnewline
$100$  &  $5$ & $500$ &  $ 0\ (0.98)$  &  $0\ (1.00)$  &  $0\ (0.97)$  &  $0\ (0.76)$  &  $0\ (0.78)$  &  $ 72.99$  &  $34.25$  &  $19.15$  &  $23.53$ \tabularnewline
$100$  &  $5$ & $1000$ & $ 1\ (0.88)$  &  $0\ (1.00)$  &  $0\ (0.97)$  &  $0\ (0.73)$  &  $0\ (0.76)$  &  $ 72.41$  &  $34.49$  &  $19.78$  &  $23.81$ \tabularnewline
\addlinespace
$200$  &  $5$ & $10$ &   $ 0\ (1.00)$  &  $0\ (1.00)$  &  $0\ (1.00)$  &  $0\ (0.89)$  &  $0\ (0.82)$  &  $259.63$  &  $76.34$  &  $44.28$  &  $55.53$ \tabularnewline
$200$  &  $5$ & $25$ &   $ 0\ (1.00)$  &  $0\ (1.00)$  &  $0\ (1.00)$  &  $0\ (0.72)$  &  $0\ (0.71)$  &  $259.76$  &  $76.57$  &  $44.01$  &  $55.23$ \tabularnewline
$200$  &  $5$ & $100$ &  $ 0\ (1.00)$  &  $0\ (1.00)$  &  $0\ (1.00)$  &  $0\ (0.73)$  &  $0\ (0.73)$  &  $274.21$  &  $77.72$  &  $45.59$  &  $56.19$ \tabularnewline
$200$  &  $5$ & $500$ &  $ 0\ (1.00)$  &  $0\ (1.00)$  &  $0\ (0.99)$  &  $1\ (0.74)$  &  $1\ (0.75)$  &  $281.71$  &  $77.21$  &  $46.14$  &  $56.43$ \tabularnewline
$200$  &  $5$ & $1000$ & $ 0\ (1.00)$  &  $0\ (1.00)$  &  $0\ (1.00)$  &  $0\ (0.73)$  &  $0\ (0.74)$  &  $275.00$  &  $76.04$  &  $45.95$  &  $55.88$ \tabularnewline
\bottomrule
\end{tabular}

	\caption{Complete results for Simulation~\ref{sim:K1}. Number of false positives at $\alpha = 0.05$, mean $p$-value, median computation time over 100 replications.}
	\label{supptable:K1results}
\end{table}
\begin{table}[h]\footnotesize
	\centering
	\begin{tabular}{ccc rrrrr rrrr}
\toprule
\multicolumn{3}{c}{parameters}&
	\multicolumn{5}{c}{$|p\text{-value}<0.05|$ (mean $p$-value)}&
	\multicolumn{4}{c}{median time (sec.)}
\tabularnewline
\cmidrule(r){1-3} \cmidrule(r){4-8} \cmidrule{9-12}
\multicolumn{1}{c}{$n_k$}&
	\multicolumn{1}{c}{$p$}&
	\multicolumn{1}{c}{$\delta$}&
	\multicolumn{1}{c}{\texttt{pv}AU}&
	\multicolumn{1}{c}{\texttt{pv}BP}&
	\multicolumn{1}{c}{SHC1}&
	\multicolumn{1}{c}{$\text{SHC2}_L$}&
	\multicolumn{1}{c}{$\text{SHC2}_2$}&
	\multicolumn{1}{c}{\texttt{pv}}&
	\multicolumn{1}{c}{SHC1}&
	\multicolumn{1}{c}{$\text{SHC2}_L$}&
	\multicolumn{1}{c}{$\text{SHC2}_2$}
\tabularnewline
\midrule
$ 50$ & $   2$ & $ 1$ &           $-$ &          $-$ &  $  0\ (0.81)$ &  $  0\ (0.55)$ &  $  0\ (0.55)$ &       $-$ &  $ 1.37$ &   $0.54$ &  $ 1.13$ \tabularnewline
$ 50$ & $   2$ & $ 2$ &           $-$ &          $-$ &  $  1\ (0.56)$ &  $  1\ (0.36)$ &   $13\ (0.32)$ &       $-$ &  $ 1.37$ &   $0.55$ &  $ 1.13$ \tabularnewline
$ 50$ & $   2$ & $ 3$ &           $-$ &          $-$ &  $ 67\ (0.10)$ &  $ 18\ (0.13)$ &   $77\ (0.04)$ &       $-$ &  $ 1.42$ &   $0.56$ &  $ 1.12$ \tabularnewline
$ 50$ & $   2$ & $ 4$ &           $-$ &          $-$ &  $ 98\ (0.00)$ &  $ 81\ (0.04)$ &   $99\ (0.00)$ &       $-$ &  $ 1.42$ &   $0.54$ &  $ 1.21$ \tabularnewline
$ 50$ & $   2$ & $ 5$ &           $-$ &          $-$ &  $100\ (0.00)$ &  $ 98\ (0.01)$ &  $100\ (0.00)$ &       $-$ &  $ 1.36$ &   $0.51$ &  $ 1.08$ \tabularnewline
\addlinespace
$100$ & $   2$ & $ 1$ &           $-$ &          $-$ &  $  0\ (0.88)$ &  $  0\ (0.56)$ &  $  0\ (0.56)$ &       $-$ &  $ 2.89$ &  $ 1.20$ &  $ 2.50$ \tabularnewline
$100$ & $   2$ & $ 2$ &           $-$ &          $-$ &  $  2\ (0.59)$ &  $  1\ (0.31)$ &  $ 14\ (0.27)$ &       $-$ &  $ 3.02$ &  $ 1.16$ &  $ 2.62$ \tabularnewline
$100$ & $   2$ & $ 3$ &           $-$ &          $-$ &  $ 80\ (0.07)$ &  $ 53\ (0.07)$ &  $ 86\ (0.03)$ &       $-$ &  $ 2.85$ &  $ 1.27$ &  $ 2.31$ \tabularnewline
$100$ & $   2$ & $ 4$ &           $-$ &          $-$ &  $100\ (0.00)$ &  $ 98\ (0.01)$ &  $100\ (0.00)$ &       $-$ &  $ 2.92$ &  $ 1.25$ &  $ 2.37$ \tabularnewline
$100$ & $   2$ & $ 5$ &           $-$ &          $-$ &  $100\ (0.00)$ &  $100\ (0.00)$ &  $100\ (0.00)$ &       $-$ &  $ 2.86$ &  $ 1.18$ &  $ 2.52$ \tabularnewline
\addlinespace
$ 50$ & $1000$ & $ 2$ &  $ 0\ (1.00)$ &  $0\ (1.00)$ &  $  0\ (1.00)$ &  $  0\ (1.00)$ &  $  0\ (1.00)$ &  $ 79.05$ &  $35.82$ &  $20.92$ &  $24.98$ \tabularnewline
$ 50$ & $1000$ & $ 4$ &  $ 0\ (1.00)$ &  $0\ (1.00)$ &  $  0\ (1.00)$ &  $  0\ (1.00)$ &  $  0\ (1.00)$ &  $ 78.60$ &  $34.99$ &  $20.49$ &  $24.73$ \tabularnewline
$ 50$ & $1000$ & $ 6$ &  $ 0\ (1.00)$ &  $0\ (1.00)$ &  $  0\ (0.99)$ &  $  0\ (1.00)$ &  $  1\ (0.81)$ &  $ 78.51$ &  $34.41$ &  $20.05$ &  $24.20$ \tabularnewline
$ 50$ & $1000$ & $ 8$ &  $ 1\ (0.99)$ &  $0\ (1.00)$ &  $ 18\ (0.23)$ &  $ 28\ (0.14)$ &   $97\ (0.01)$ &  $ 78.51$ &  $34.95$ &  $20.51$ &  $24.41$ \tabularnewline
$ 50$ & $1000$ & $10$ &  $ 1\ (0.68)$ &  $0\ (0.92)$ &  $ 99\ (0.00)$ &  $100\ (0.00)$ &  $100\ (0.00)$ &  $ 78.76$ &  $35.35$ &  $20.62$ &  $24.57$ \tabularnewline
$ 50$ & $1000$ & $12$ &  $ 1\ (0.50)$ &  $0\ (0.68)$ &  $100\ (0.00)$ &  $100\ (0.00)$ &  $100\ (0.00)$ &  $ 78.68$ &  $35.03$ &  $20.51$ &  $24.64$ \tabularnewline
$ 50$ & $1000$ & $14$ &  $ 6\ (0.25)$ &  $0\ (0.52)$ &  $100\ (0.00)$ &  $100\ (0.00)$ &  $100\ (0.00)$ &  $ 78.36$ &  $35.04$ &  $19.92$ &  $24.41$ \tabularnewline
$ 50$ & $1000$ & $16$ &  $48\ (0.11)$ &  $0\ (0.42)$ &  $100\ (0.00)$ &  $100\ (0.00)$ &  $100\ (0.00)$ &  $ 77.75$ &  $34.63$ &  $19.76$ &  $24.12$ \tabularnewline
$ 50$ & $1000$ & $18$ &  $75\ (0.07)$ &  $0\ (0.39)$ &  $100\ (0.00)$ &  $100\ (0.00)$ &  $100\ (0.00)$ &  $ 75.94$ &  $34.79$ &  $19.69$ &  $24.11$ \tabularnewline
$ 50$ & $1000$ & $20$ &  $84\ (0.12)$ &  $0\ (0.42)$ &  $100\ (0.00)$ &  $100\ (0.00)$ &  $100\ (0.00)$ &  $ 74.95$ &  $34.55$ &  $19.61$ &  $23.76$ \tabularnewline
\addlinespace
$100$ & $1000$ & $ 2$ &  $ 0\ (1.00)$ &  $0\ (1.00)$ &  $  0\ (1.00)$ &  $  0\ (1.00)$ &  $  0\ (1.00)$ &  $342.34$ &  $85.62$ &  $51.24$ &  $62.91$ \tabularnewline
$100$ & $1000$ & $ 4$ &  $ 0\ (1.00)$ &  $0\ (1.00)$ &  $  0\ (1.00)$ &  $  0\ (1.00)$ &  $  0\ (1.00)$ &  $329.18$ &  $86.59$ &  $50.56$ &  $62.71$ \tabularnewline
$100$ & $1000$ & $ 6$ &  $ 0\ (1.00)$ &  $0\ (1.00)$ &  $  0\ (0.95)$ &  $ 21\ (0.29)$ &  $ 57\ (0.10)$ &  $320.53$ &  $86.29$ &  $49.44$ &  $62.58$ \tabularnewline
$100$ & $1000$ & $ 8$ &  $ 0\ (1.00)$ &  $0\ (1.00)$ &  $ 84\ (0.03)$ &  $100\ (0.00)$ &  $100\ (0.00)$ &  $311.57$ &  $86.83$ &  $51.29$ &  $65.07$ \tabularnewline
$100$ & $1000$ & $10$ &  $ 0\ (0.81)$ &  $0\ (0.97)$ &  $100\ (0.00)$ &  $100\ (0.00)$ &  $100\ (0.00)$ &  $314.98$ &  $90.41$ &  $50.89$ &  $63.54$ \tabularnewline
$100$ & $1000$ & $12$ &  $ 1\ (0.58)$ &  $0\ (0.75)$ &  $100\ (0.00)$ &  $100\ (0.00)$ &  $100\ (0.00)$ &  $321.48$ &  $85.76$ &  $50.40$ &  $60.86$ \tabularnewline
$100$ & $1000$ & $14$ &  $ 0\ (0.29)$ &  $0\ (0.51)$ &  $100\ (0.00)$ &  $100\ (0.00)$ &  $100\ (0.00)$ &  $314.38$ &  $86.93$ &  $50.78$ &  $63.36$ \tabularnewline
$100$ & $1000$ & $16$ &  $43\ (0.15)$ &  $0\ (0.44)$ &  $100\ (0.00)$ &  $100\ (0.00)$ &  $100\ (0.00)$ &  $311.31$ &  $84.92$ &  $50.30$ &  $61.36$ \tabularnewline
$100$ & $1000$ & $18$ &  $78\ (0.10)$ &  $0\ (0.41)$ &  $100\ (0.00)$ &  $100\ (0.00)$ &  $100\ (0.00)$ &  $315.93$ &  $84.87$ &  $50.21$ &  $63.05$ \tabularnewline
$100$ & $1000$ & $20$ &  $89\ (0.09)$ &  $0\ (0.40)$ &  $100\ (0.00)$ &  $100\ (0.00)$ &  $100\ (0.00)$ &  $295.15$ &  $81.43$ &  $48.60$ &  $59.26$ \tabularnewline
\bottomrule
\end{tabular}
	\caption{Complete results for the $K=2$ alternative setting. Number of replications identifying the correct number of significant clusters, mean $p$-value, median computation time over 100 replications.}
	\label{supptable:K2results}
\end{table}
\begin{table}[t]\footnotesize
	\centering
	\begin{tabular}{ccc rrrrr rrrr}
\toprule
\multicolumn{3}{c}{parameters}&
	\multicolumn{5}{c}{$|\hat{K}=3|$}&
	\multicolumn{4}{c}{median time (sec.)}
\tabularnewline
\cmidrule(r){1-3} \cmidrule(r){4-8} \cmidrule(r){9-12}
\multicolumn{1}{c}{$n_k$}&
	\multicolumn{1}{c}{$p$}&
	\multicolumn{1}{c}{$\delta$}&
	\multicolumn{1}{c}{\texttt{pv}AU}&
	\multicolumn{1}{c}{\texttt{pv}BP}&
	\multicolumn{1}{c}{SHC1}&
	\multicolumn{1}{c}{$\text{SHC2}_L$}&
	\multicolumn{1}{c}{$\text{SHC2}_2$}&
	\multicolumn{1}{c}{\texttt{pv}}&
	\multicolumn{1}{c}{SHC1}&
	\multicolumn{1}{c}{$\text{SHC2}_L$}&
	\multicolumn{1}{c}{$\text{SHC2}_2$}
\tabularnewline
\midrule
$ 50$&$   2$&$ 1$&$-$&$-$&$  0$&$  0$&$  0$&$-$&$  2.22$&$  0.99$&$  1.77$\tabularnewline
$ 50$&$   2$&$ 2$&$-$&$-$&$  0$&$  0$&$  1$&$-$&$  2.17$&$  0.98$&$  1.73$\tabularnewline
$ 50$&$   2$&$ 3$&$-$&$-$&$ 20$&$  0$&$ 30$&$-$&$  2.23$&$  0.99$&$  1.77$\tabularnewline
$ 50$&$   2$&$ 4$&$-$&$-$&$ 79$&$  3$&$ 85$&$-$&$  2.24$&$  0.99$&$  1.80$\tabularnewline
$ 50$&$   2$&$ 5$&$-$&$-$&$ 98$&$  9$&$ 99$&$-$&$  2.23$&$  1.00$&$  1.78$\tabularnewline
\addlinespace
$100$&$   2$&$ 1$&$-$&$-$&$  0$&$  0$&$  0$&$-$&$  4.88$&$  2.41$&$  3.92$\tabularnewline
$100$&$   2$&$ 2$&$-$&$-$&$  0$&$  0$&$  3$&$-$&$  4.78$&$  2.37$&$  3.85$\tabularnewline
$100$&$   2$&$ 3$&$-$&$-$&$ 46$&$  8$&$ 60$&$-$&$  4.85$&$  2.40$&$  3.92$\tabularnewline
$100$&$   2$&$ 4$&$-$&$-$&$ 93$&$ 44$&$ 96$&$-$&$  4.87$&$  2.42$&$  3.93$\tabularnewline
$100$&$   2$&$ 5$&$-$&$-$&$100$&$ 72$&$100$&$-$&$  4.90$&$  2.42$&$  3.95$\tabularnewline
\addlinespace
$ 50$&$1000$&$ 2$&$ 0$&$ 0$&$  0$&$  0$&$  0$&$200.13$&$ 72.28$&$ 41.60$&$ 49.02$\tabularnewline
$ 50$&$1000$&$ 4$&$ 0$&$ 0$&$  0$&$  0$&$  0$&$199.97$&$ 62.44$&$ 38.41$&$ 45.68$\tabularnewline
$ 50$&$1000$&$ 6$&$ 0$&$ 0$&$  0$&$  0$&$  0$&$200.07$&$ 63.92$&$ 40.08$&$ 47.30$\tabularnewline
$ 50$&$1000$&$ 8$&$ 0$&$ 0$&$  0$&$  2$&$ 69$&$199.51$&$ 63.79$&$ 39.84$&$ 47.00$\tabularnewline
$ 50$&$1000$&$10$&$ 0$&$ 0$&$  5$&$ 91$&$ 97$&$199.88$&$ 66.00$&$ 40.53$&$ 48.15$\tabularnewline
$ 50$&$1000$&$12$&$ 0$&$ 0$&$ 19$&$ 93$&$100$&$199.34$&$ 65.57$&$ 41.28$&$ 48.58$\tabularnewline
$ 50$&$1000$&$14$&$ 0$&$ 0$&$ 20$&$ 87$&$100$&$199.98$&$ 65.49$&$ 41.16$&$ 48.18$\tabularnewline
$ 50$&$1000$&$16$&$ 3$&$ 0$&$ 48$&$ 84$&$ 98$&$199.62$&$ 66.19$&$ 40.76$&$ 48.14$\tabularnewline
$ 50$&$1000$&$18$&$ 9$&$ 0$&$ 56$&$ 81$&$100$&$199.78$&$ 65.33$&$ 40.78$&$ 48.09$\tabularnewline
$ 50$&$1000$&$20$&$ 8$&$ 0$&$ 73$&$ 71$&$100$&$200.12$&$ 65.53$&$ 41.47$&$ 48.91$\tabularnewline
\addlinespace
$100$&$1000$&$ 2$&$ 0$&$ 0$&$  0$&$  0$&$  0$&$762.83$&$155.28$&$101.07$&$120.10$\tabularnewline
$100$&$1000$&$ 4$&$ 0$&$ 0$&$  0$&$  0$&$  0$&$763.74$&$154.52$&$ 97.56$&$116.54$\tabularnewline
$100$&$1000$&$ 6$&$ 0$&$ 0$&$  0$&$  2$&$ 26$&$768.71$&$153.66$&$ 98.31$&$117.24$\tabularnewline
$100$&$1000$&$ 8$&$ 0$&$ 0$&$  1$&$ 99$&$ 98$&$768.34$&$160.74$&$101.88$&$119.63$\tabularnewline
$100$&$1000$&$10$&$ 0$&$ 0$&$ 71$&$100$&$100$&$775.47$&$176.17$&$101.00$&$131.41$\tabularnewline
$100$&$1000$&$12$&$ 0$&$ 0$&$ 98$&$100$&$100$&$881.97$&$153.13$&$ 99.08$&$118.12$\tabularnewline
$100$&$1000$&$14$&$ 0$&$ 0$&$ 99$&$100$&$100$&$882.03$&$175.52$&$113.72$&$121.56$\tabularnewline
$100$&$1000$&$16$&$ 1$&$ 0$&$100$&$100$&$100$&$883.02$&$152.94$&$ 97.92$&$135.61$\tabularnewline
$100$&$1000$&$18$&$12$&$ 0$&$100$&$100$&$100$&$883.94$&$180.11$&$100.53$&$120.02$\tabularnewline
$100$&$1000$&$20$&$ 6$&$ 0$&$100$&$100$&$100$&$881.96$&$153.88$&$113.75$&$118.49$\tabularnewline
\bottomrule
\end{tabular}

	\caption{Complete results for the ``line" arrangement considered in Simulation~\ref{sim:K3}. Number of replications identifying the correct number of significant clusters, median computation time over 100 replications.}
	\label{supptable:K3results1}
\end{table}
\begin{table}[t]\footnotesize
	\centering
	\begin{tabular}{ccc rrrrr rrrr}
\toprule
\multicolumn{3}{c}{parameters}&
	\multicolumn{5}{c}{$|\hat{K}=3|$}&
	\multicolumn{4}{c}{median time (sec.)}
\tabularnewline
\cmidrule(r){1-3} \cmidrule(r){4-8} \cmidrule(r){9-12}
\multicolumn{1}{c}{$n_k$}&
	\multicolumn{1}{c}{$p$}&
	\multicolumn{1}{c}{$\delta$}&
	\multicolumn{1}{c}{\texttt{pv}AU}&
	\multicolumn{1}{c}{\texttt{pv}BP}&
	\multicolumn{1}{c}{SHC1}&
	\multicolumn{1}{c}{$\text{SHC2}_L$}&
	\multicolumn{1}{c}{$\text{SHC2}_2$}&
	\multicolumn{1}{c}{\texttt{pv}}&
	\multicolumn{1}{c}{SHC1}&
	\multicolumn{1}{c}{$\text{SHC2}_L$}&
	\multicolumn{1}{c}{$\text{SHC2}_2$}
\tabularnewline
\midrule
$ 50$&$   2$&$ 1$&$-$&$-$&$  0$&$  0$&$  0$&$-$&$  2.22$&$  1.00$&$  1.77$\tabularnewline
$ 50$&$   2$&$ 2$&$-$&$-$&$  0$&$  0$&$  0$&$-$&$  2.13$&$  0.98$&$  1.76$\tabularnewline
$ 50$&$   2$&$ 3$&$-$&$-$&$  0$&$  0$&$  8$&$-$&$  2.22$&$  1.00$&$  1.78$\tabularnewline
$ 50$&$   2$&$ 4$&$-$&$-$&$ 28$&$ 29$&$ 81$&$-$&$  2.18$&$  0.98$&$  1.76$\tabularnewline
$ 50$&$   2$&$ 5$&$-$&$-$&$ 98$&$ 94$&$ 99$&$-$&$  2.22$&$  1.00$&$  1.76$\tabularnewline
\addlinespace
$100$&$   2$&$ 1$&$-$&$-$&$  0$&$  0$&$  0$&$-$&$  4.86$&$  2.44$&$  3.94$\tabularnewline
$100$&$   2$&$ 2$&$-$&$-$&$  0$&$  0$&$  0$&$-$&$  4.84$&$  2.40$&$  3.94$\tabularnewline
$100$&$   2$&$ 3$&$-$&$-$&$  2$&$ 11$&$ 32$&$-$&$  4.81$&$  2.40$&$  3.89$\tabularnewline
$100$&$   2$&$ 4$&$-$&$-$&$ 72$&$ 89$&$100$&$-$&$  4.93$&$  2.44$&$  3.93$\tabularnewline
$100$&$   2$&$ 5$&$-$&$-$&$100$&$100$&$100$&$-$&$  4.86$&$  2.43$&$  3.92$\tabularnewline
\addlinespace
$ 50$&$1000$&$ 2$&$ 0$&$ 0$&$  0$&$  0$&$  0$&$232.70$&$ 75.08$&$ 46.81$&$ 56.06$\tabularnewline
$ 50$&$1000$&$ 4$&$ 0$&$ 0$&$  0$&$  0$&$  0$&$232.78$&$ 75.45$&$ 47.00$&$ 56.05$\tabularnewline
$ 50$&$1000$&$ 6$&$ 0$&$ 0$&$  0$&$  0$&$  0$&$232.53$&$ 76.34$&$ 47.24$&$ 56.53$\tabularnewline
$ 50$&$1000$&$ 8$&$ 0$&$ 0$&$  0$&$  1$&$ 78$&$232.41$&$ 75.60$&$ 46.97$&$ 56.26$\tabularnewline
$ 50$&$1000$&$10$&$ 0$&$ 0$&$ 89$&$100$&$100$&$232.28$&$ 76.72$&$ 47.90$&$ 57.14$\tabularnewline
$ 50$&$1000$&$12$&$ 0$&$ 0$&$100$&$100$&$100$&$232.50$&$ 76.51$&$ 47.84$&$ 57.37$\tabularnewline
$ 50$&$1000$&$14$&$ 0$&$ 0$&$100$&$100$&$100$&$232.32$&$ 75.86$&$ 47.16$&$ 56.51$\tabularnewline
$ 50$&$1000$&$16$&$12$&$ 0$&$100$&$100$&$100$&$232.47$&$ 75.57$&$ 47.38$&$ 56.62$\tabularnewline
$ 50$&$1000$&$18$&$48$&$ 0$&$100$&$100$&$100$&$232.28$&$ 76.29$&$ 47.43$&$ 57.08$\tabularnewline
$ 50$&$1000$&$20$&$33$&$ 0$&$100$&$100$&$100$&$232.46$&$ 75.72$&$ 47.29$&$ 56.83$\tabularnewline
\addlinespace
$100$&$1000$&$ 2$&$ 0$&$ 0$&$  0$&$  0$&$  0$&$885.44$&$176.23$&$113.44$&$137.73$\tabularnewline
$100$&$1000$&$ 4$&$ 0$&$ 0$&$  0$&$  0$&$  0$&$885.49$&$179.78$&$115.81$&$140.23$\tabularnewline
$100$&$1000$&$ 6$&$ 0$&$ 0$&$  0$&$  0$&$  5$&$885.93$&$181.16$&$116.15$&$140.19$\tabularnewline
$100$&$1000$&$ 8$&$ 0$&$ 0$&$  2$&$100$&$100$&$885.79$&$182.45$&$117.99$&$142.00$\tabularnewline
$100$&$1000$&$10$&$ 0$&$ 0$&$100$&$100$&$100$&$885.94$&$183.59$&$118.18$&$142.90$\tabularnewline
$100$&$1000$&$12$&$ 0$&$ 0$&$100$&$100$&$100$&$885.36$&$181.59$&$116.98$&$141.46$\tabularnewline
$100$&$1000$&$14$&$ 0$&$ 0$&$100$&$100$&$100$&$886.11$&$184.46$&$117.54$&$141.82$\tabularnewline
$100$&$1000$&$16$&$ 4$&$ 0$&$100$&$100$&$100$&$886.11$&$182.38$&$117.42$&$142.67$\tabularnewline
$100$&$1000$&$18$&$39$&$ 0$&$100$&$100$&$100$&$886.50$&$181.53$&$116.95$&$140.76$\tabularnewline
$100$&$1000$&$20$&$47$&$ 0$&$100$&$100$&$100$&$886.73$&$179.93$&$116.39$&$140.08$\tabularnewline
\bottomrule
\end{tabular}
	\caption{Complete results for the ``triangle" arrangement considered in Simulation~\ref{sim:K3}. Number of replications identifying the correct number of significant clusters, median computation time over 100 replications.}
	\label{supptable:K3results2}
\end{table}
\begin{table}[t]\footnotesize
	\centering
	\begin{tabular}{ccc rrrrr rrrr}
\toprule
\multicolumn{3}{c}{parameters}&
	\multicolumn{5}{c}{$|\hat{K}=4|$}&
	\multicolumn{4}{c}{median time (sec.)}
\tabularnewline
\cmidrule(r){1-3} \cmidrule(r){4-8} \cmidrule(r){9-12}
\multicolumn{1}{c}{$n_k$}&
	\multicolumn{1}{c}{$p$}&
	\multicolumn{1}{c}{$\delta$}&
	\multicolumn{1}{c}{\texttt{pv}AU}&
	\multicolumn{1}{c}{\texttt{pv}BP}&
	\multicolumn{1}{c}{SHC1}&
	\multicolumn{1}{c}{$\text{SHC2}_L$}&
	\multicolumn{1}{c}{$\text{SHC2}_2$}&
	\multicolumn{1}{c}{\texttt{pv}}&
	\multicolumn{1}{c}{SHC1}&
	\multicolumn{1}{c}{$\text{SHC2}_L$}&
	\multicolumn{1}{c}{$\text{SHC2}_2$}
\tabularnewline
\midrule
$ 50$&$   2$&$ 1$&$-$&$-$&$  0$&$  0$&$  0$&$-$&$  2.47$&$ 1.09$&$ 2.08$\tabularnewline
$ 50$&$   2$&$ 2$&$-$&$-$&$  0$&$  0$&$  0$&$-$&$  2.29$&$ 1.04$&$ 1.95$\tabularnewline
$ 50$&$   2$&$ 3$&$-$&$-$&$  3$&$  0$&$ 17$&$-$&$  2.54$&$ 1.16$&$ 2.06$\tabularnewline
$ 50$&$   2$&$ 4$&$-$&$-$&$ 78$&$ 12$&$ 90$&$-$&$  2.57$&$ 1.16$&$ 2.07$\tabularnewline
$ 50$&$   2$&$ 5$&$-$&$-$&$100$&$ 84$&$100$&$-$&$  2.70$&$ 1.22$&$ 2.16$\tabularnewline
\addlinespace
$100$&$   2$&$ 1$&$-$&$-$&$  0$&$  0$&$  0$&$-$&$  5.31$&$ 2.69$&$ 4.61$\tabularnewline
$100$&$   2$&$ 2$&$-$&$-$&$  0$&$  0$&$  0$&$-$&$  5.14$&$ 2.60$&$ 4.32$\tabularnewline
$100$&$   2$&$ 3$&$-$&$-$&$ 18$&$  4$&$ 53$&$-$&$  5.33$&$ 2.54$&$ 4.42$\tabularnewline
$100$&$   2$&$ 4$&$-$&$-$&$ 98$&$ 85$&$ 99$&$-$&$  4.87$&$ 2.43$&$ 4.11$\tabularnewline
$100$&$   2$&$ 5$&$-$&$-$&$100$&$100$&$100$&$-$&$  5.04$&$ 2.37$&$ 4.39$\tabularnewline
\addlinespace
$ 50$&$1000$&$ 2$&$ 0$&$ 0$&$  0$&$  0$&$  0$&$305.45$&$ 81.27$&$49.64$&$59.28$\tabularnewline
$ 50$&$1000$&$ 4$&$ 0$&$ 0$&$  0$&$  0$&$  0$&$299.88$&$ 80.44$&$48.94$&$58.58$\tabularnewline
$ 50$&$1000$&$ 6$&$ 0$&$ 0$&$  0$&$  0$&$  0$&$299.85$&$ 80.59$&$48.69$&$57.49$\tabularnewline
$ 50$&$1000$&$ 8$&$ 0$&$ 0$&$  0$&$  1$&$ 67$&$300.88$&$ 81.43$&$49.38$&$58.52$\tabularnewline
$ 50$&$1000$&$10$&$ 0$&$ 0$&$ 95$&$100$&$100$&$301.08$&$ 82.16$&$50.41$&$59.38$\tabularnewline
$ 50$&$1000$&$12$&$ 0$&$ 0$&$100$&$100$&$100$&$300.91$&$ 82.17$&$49.69$&$59.50$\tabularnewline
$ 50$&$1000$&$14$&$ 1$&$ 0$&$100$&$100$&$100$&$298.67$&$ 81.54$&$49.50$&$58.98$\tabularnewline
$ 50$&$1000$&$16$&$77$&$ 0$&$100$&$100$&$100$&$405.38$&$109.32$&$68.26$&$81.71$\tabularnewline
$ 50$&$1000$&$18$&$97$&$ 0$&$100$&$100$&$100$&$402.78$&$109.64$&$68.78$&$82.11$\tabularnewline
$ 50$&$1000$&$20$&$99$&$ 0$&$100$&$100$&$100$&$403.52$&$110.56$&$68.81$&$82.55$\tabularnewline
\bottomrule
\end{tabular}

	\caption{Complete results for the ``square" arrangement considered in Simulation~\ref{sim:K4}. Number of replications identifying the correct number of significant clusters, median computation time over 100 replications.}
	\label{supptable:K4results1}
\end{table}
\begin{table}[t]\footnotesize
	\centering
	\begin{tabular}{ccc rrrrr rrrr}
\toprule
\multicolumn{3}{c}{parameters}&
	\multicolumn{5}{c}{$|\hat{K}=4|$}&
	\multicolumn{4}{c}{median time (sec.)}
\tabularnewline
\cmidrule(r){1-3} \cmidrule(r){4-8} \cmidrule(r){9-12}
\multicolumn{1}{c}{$n_k$}&
	\multicolumn{1}{c}{$p$}&
	\multicolumn{1}{c}{$\delta$}&
	\multicolumn{1}{c}{\texttt{pv}AU}&
	\multicolumn{1}{c}{\texttt{pv}BP}&
	\multicolumn{1}{c}{SHC1}&
	\multicolumn{1}{c}{$\text{SHC2}_L$}&
	\multicolumn{1}{c}{$\text{SHC2}_2$}&
	\multicolumn{1}{c}{\texttt{pv}}&
	\multicolumn{1}{c}{SHC1}&
	\multicolumn{1}{c}{$\text{SHC2}_L$}&
	\multicolumn{1}{c}{$\text{SHC2}_2$}
\tabularnewline
\midrule
$ 50$&$   3$&$ 1$&$-$&$-$&$  0$&$  0$&$  0$&$-$&$  2.54$&$ 1.10$&$ 1.98$\tabularnewline
$ 50$&$   3$&$ 2$&$-$&$-$&$  0$&$  0$&$  0$&$-$&$  2.57$&$ 1.12$&$ 2.02$\tabularnewline
$ 50$&$   3$&$ 3$&$-$&$-$&$  0$&$  0$&$  0$&$-$&$  2.43$&$ 1.10$&$ 1.91$\tabularnewline
$ 50$&$   3$&$ 4$&$-$&$-$&$  0$&$  9$&$ 33$&$-$&$  2.84$&$ 1.28$&$ 2.25$\tabularnewline
$ 50$&$   3$&$ 5$&$-$&$-$&$ 24$&$ 86$&$ 99$&$-$&$  2.40$&$ 1.09$&$ 2.03$\tabularnewline
\addlinespace
$100$&$   3$&$ 1$&$-$&$-$&$  0$&$  0$&$  0$&$-$&$  5.24$&$ 2.61$&$ 4.44$\tabularnewline
$100$&$   3$&$ 2$&$-$&$-$&$  0$&$  0$&$  0$&$-$&$  5.79$&$ 2.74$&$ 4.77$\tabularnewline
$100$&$   3$&$ 3$&$-$&$-$&$  0$&$  1$&$  2$&$-$&$  5.80$&$ 2.95$&$ 4.55$\tabularnewline
$100$&$   3$&$ 4$&$-$&$-$&$  2$&$ 84$&$ 94$&$-$&$  5.99$&$ 2.92$&$ 4.76$\tabularnewline
$100$&$   3$&$ 5$&$-$&$-$&$ 88$&$ 99$&$100$&$-$&$  5.11$&$ 2.63$&$ 4.19$\tabularnewline
\addlinespace
$ 50$&$1000$&$ 2$&$ 0$&$ 0$&$  0$&$  0$&$  0$&$370.14$&$ 94.85$&$58.06$&$69.12$\tabularnewline
$ 50$&$1000$&$ 4$&$ 0$&$ 0$&$  0$&$  0$&$  0$&$363.98$&$ 92.81$&$57.70$&$69.67$\tabularnewline
$ 50$&$1000$&$ 6$&$ 0$&$ 0$&$  0$&$  0$&$  0$&$367.20$&$ 95.37$&$57.39$&$75.23$\tabularnewline
$ 50$&$1000$&$ 8$&$ 0$&$ 0$&$  0$&$  0$&$ 37$&$385.62$&$ 93.53$&$58.15$&$71.01$\tabularnewline
$ 50$&$1000$&$10$&$ 0$&$ 0$&$ 56$&$ 98$&$100$&$364.72$&$ 98.81$&$60.74$&$72.49$\tabularnewline
$ 50$&$1000$&$12$&$ 0$&$ 0$&$100$&$100$&$100$&$383.10$&$ 98.10$&$61.74$&$78.22$\tabularnewline
$ 50$&$1000$&$14$&$ 0$&$ 0$&$100$&$100$&$100$&$368.79$&$ 98.68$&$60.75$&$77.42$\tabularnewline
$ 50$&$1000$&$16$&$16$&$ 0$&$100$&$100$&$100$&$403.31$&$114.20$&$71.40$&$85.96$\tabularnewline
$ 50$&$1000$&$18$&$53$&$ 0$&$100$&$100$&$100$&$402.61$&$112.10$&$70.47$&$84.58$\tabularnewline
$ 50$&$1000$&$20$&$68$&$ 0$&$100$&$100$&$100$&$404.02$&$113.07$&$70.92$&$85.09$\tabularnewline
\bottomrule
\end{tabular}

	\caption{Complete results for the ``tetrahedron" arrangement considered in Simulation~\ref{sim:K4}. Number of replications identifying the correct number of significant clusters, median computation time over 100 replications.}
	\label{supptable:K4results2}
\end{table}
\begin{table}[t]\footnotesize
	\centering
	\begin{tabular}{ccc rrrrr rrrr}
\toprule
\multicolumn{3}{c}{parameters}&
	\multicolumn{5}{c}{$|\hat{K}=4|$}&
	\multicolumn{4}{c}{median time (sec.)}
\tabularnewline
\cmidrule(r){1-3} \cmidrule(r){4-8} \cmidrule(r){9-12}
\multicolumn{1}{c}{$n_k$}&
	\multicolumn{1}{c}{$p$}&
	\multicolumn{1}{c}{$\delta$}&
	\multicolumn{1}{c}{\texttt{pv}AU}&
	\multicolumn{1}{c}{\texttt{pv}BP}&
	\multicolumn{1}{c}{SHC1}&
	\multicolumn{1}{c}{$\text{SHC2}_L$}&
	\multicolumn{1}{c}{$\text{SHC2}_2$}&
	\multicolumn{1}{c}{\texttt{pv}}&
	\multicolumn{1}{c}{SHC1}&
	\multicolumn{1}{c}{$\text{SHC2}_L$}&
	\multicolumn{1}{c}{$\text{SHC2}_2$}
\tabularnewline
\midrule
$ 50$&$   2$&$ 1$&$-$&$-$&$  0$&$  0$&$  0$&$-$&$  2.19$&$ 0.99$&$ 1.76$\tabularnewline
$ 50$&$   2$&$ 2$&$-$&$-$&$  0$&$  0$&$  0$&$-$&$  2.20$&$ 0.97$&$ 1.78$\tabularnewline
$ 50$&$   2$&$ 3$&$-$&$-$&$ 10$&$  0$&$ 22$&$-$&$  2.20$&$ 0.99$&$ 1.75$\tabularnewline
$ 50$&$   2$&$ 4$&$-$&$-$&$ 88$&$ 26$&$ 94$&$-$&$  2.49$&$ 1.11$&$ 1.95$\tabularnewline
$ 50$&$   2$&$ 5$&$-$&$-$&$100$&$ 96$&$100$&$-$&$  2.49$&$ 1.07$&$ 1.97$\tabularnewline
\addlinespace
$100$&$   2$&$ 1$&$-$&$-$&$  0$&$  0$&$  0$&$-$&$  5.12$&$ 2.66$&$ 4.30$\tabularnewline
$100$&$   2$&$ 2$&$-$&$-$&$  0$&$  0$&$  0$&$-$&$  5.03$&$ 2.58$&$ 4.12$\tabularnewline
$100$&$   2$&$ 3$&$-$&$-$&$ 43$&$  7$&$ 54$&$-$&$  4.77$&$ 2.36$&$ 3.98$\tabularnewline
$100$&$   2$&$ 4$&$-$&$-$&$ 98$&$ 89$&$ 99$&$-$&$  4.78$&$ 2.38$&$ 3.98$\tabularnewline
$100$&$   2$&$ 5$&$-$&$-$&$100$&$100$&$100$&$-$&$  4.80$&$ 2.37$&$ 3.93$\tabularnewline
\addlinespace
$ 50$&$1000$&$ 2$&$ 0$&$ 0$&$  0$&$  0$&$  0$&$402.83$&$108.12$&$67.98$&$81.53$\tabularnewline
$ 50$&$1000$&$ 4$&$ 0$&$ 0$&$  0$&$  0$&$  0$&$402.46$&$107.61$&$67.66$&$81.23$\tabularnewline
$ 50$&$1000$&$ 6$&$ 0$&$ 0$&$  0$&$  0$&$  0$&$402.53$&$108.08$&$67.88$&$81.06$\tabularnewline
$ 50$&$1000$&$ 8$&$ 0$&$ 0$&$  0$&$  1$&$ 78$&$401.98$&$109.30$&$68.57$&$81.89$\tabularnewline
$ 50$&$1000$&$10$&$ 0$&$ 0$&$ 98$&$ 99$&$100$&$401.32$&$107.70$&$67.78$&$81.17$\tabularnewline
$ 50$&$1000$&$12$&$ 0$&$ 0$&$100$&$100$&$100$&$401.78$&$108.67$&$68.35$&$81.28$\tabularnewline
$ 50$&$1000$&$14$&$22$&$ 0$&$100$&$100$&$100$&$401.97$&$108.83$&$68.82$&$81.95$\tabularnewline
$ 50$&$1000$&$16$&$58$&$ 0$&$100$&$100$&$100$&$401.86$&$108.93$&$68.57$&$81.90$\tabularnewline
$ 50$&$1000$&$18$&$66$&$ 0$&$100$&$100$&$100$&$401.78$&$107.34$&$67.93$&$80.97$\tabularnewline
$ 50$&$1000$&$20$&$64$&$ 0$&$100$&$100$&$100$&$401.13$&$108.17$&$68.24$&$81.58$\tabularnewline
\bottomrule
\end{tabular}

	\caption{Complete results for the ``rectangle" arrangement considered in Simulation~\ref{sim:K4}. Number of replications identifying the correct number of significant clusters, median computation time over 100 replications.}
	\label{supptable:K4results3}
\end{table}
\begin{table}[t]\footnotesize
	\centering
	\begin{tabular}{ccc rrrrr rrrr}
\toprule
\multicolumn{3}{c}{parameters}&
	\multicolumn{5}{c}{$|\hat{K}=4|$}&
	\multicolumn{4}{c}{median time (sec.)}
\tabularnewline
\cmidrule(r){1-3} \cmidrule(r){4-8} \cmidrule(r){9-12}
\multicolumn{1}{c}{$n_k$}&
	\multicolumn{1}{c}{$p$}&
	\multicolumn{1}{c}{$\delta$}&
	\multicolumn{1}{c}{\texttt{pv}AU}&
	\multicolumn{1}{c}{\texttt{pv}BP}&
	\multicolumn{1}{c}{SHC1}&
	\multicolumn{1}{c}{$\text{SHC2}_L$}&
	\multicolumn{1}{c}{$\text{SHC2}_2$}&
	\multicolumn{1}{c}{\texttt{pv}}&
	\multicolumn{1}{c}{SHC1}&
	\multicolumn{1}{c}{$\text{SHC2}_L$}&
	\multicolumn{1}{c}{$\text{SHC2}_2$}
\tabularnewline
\midrule
$ 50$&$   3$&$ 1$&$-$&$-$&$  0$&$  0$&$  0$&$-$&$  2.30$&$ 1.04$&$ 1.83$\tabularnewline
$ 50$&$   3$&$ 2$&$-$&$-$&$  0$&$  0$&$  0$&$-$&$  2.41$&$ 1.09$&$ 1.95$\tabularnewline
$ 50$&$   3$&$ 3$&$-$&$-$&$  0$&$  0$&$  5$&$-$&$  2.38$&$ 1.07$&$ 1.91$\tabularnewline
$ 50$&$   3$&$ 4$&$-$&$-$&$  8$&$ 12$&$ 72$&$-$&$  2.33$&$ 1.06$&$ 1.89$\tabularnewline
$ 50$&$   3$&$ 5$&$-$&$-$&$ 88$&$ 96$&$100$&$-$&$  2.45$&$ 1.09$&$ 1.93$\tabularnewline
\addlinespace
$100$&$   3$&$ 1$&$-$&$-$&$  0$&$  0$&$  0$&$-$&$  4.96$&$ 2.54$&$ 4.14$\tabularnewline
$100$&$   3$&$ 2$&$-$&$-$&$  0$&$  0$&$  0$&$-$&$  5.04$&$ 2.59$&$ 4.13$\tabularnewline
$100$&$   3$&$ 3$&$-$&$-$&$  0$&$  9$&$ 29$&$-$&$  5.13$&$ 2.65$&$ 4.22$\tabularnewline
$100$&$   3$&$ 4$&$-$&$-$&$ 40$&$ 90$&$ 98$&$-$&$  5.23$&$ 2.63$&$ 4.21$\tabularnewline
$100$&$   3$&$ 5$&$-$&$-$&$100$&$ 99$&$100$&$-$&$  5.06$&$ 2.61$&$ 4.15$\tabularnewline
\addlinespace
$ 50$&$1000$&$ 2$&$ 0$&$ 0$&$  0$&$  0$&$  0$&$300.42$&$ 79.91$&$48.80$&$57.90$\tabularnewline
$ 50$&$1000$&$ 4$&$ 0$&$ 0$&$  0$&$  0$&$  0$&$298.10$&$ 80.43$&$49.44$&$58.76$\tabularnewline
$ 50$&$1000$&$ 6$&$ 0$&$ 0$&$  0$&$  0$&$  0$&$297.97$&$ 81.65$&$49.68$&$59.07$\tabularnewline
$ 50$&$1000$&$ 8$&$ 0$&$ 0$&$  0$&$  2$&$ 62$&$301.60$&$ 82.63$&$50.71$&$60.76$\tabularnewline
$ 50$&$1000$&$10$&$ 0$&$ 0$&$ 87$&$ 99$&$100$&$300.92$&$ 83.61$&$51.22$&$61.01$\tabularnewline
$ 50$&$1000$&$12$&$ 0$&$ 0$&$100$&$100$&$100$&$298.19$&$ 82.12$&$50.50$&$59.58$\tabularnewline
$ 50$&$1000$&$14$&$ 3$&$ 0$&$100$&$100$&$100$&$401.31$&$111.78$&$70.41$&$84.39$\tabularnewline
$ 50$&$1000$&$16$&$ 6$&$ 0$&$100$&$100$&$100$&$401.12$&$111.20$&$70.38$&$84.10$\tabularnewline
$ 50$&$1000$&$18$&$ 2$&$ 0$&$100$&$100$&$100$&$401.11$&$110.89$&$70.20$&$84.63$\tabularnewline
$ 50$&$1000$&$20$&$ 0$&$ 0$&$100$&$100$&$100$&$403.34$&$114.01$&$71.74$&$86.03$\tabularnewline
\bottomrule
\end{tabular}

	\caption{Complete results for the ``stretched tetrahedron" arrangement considered in Simulation~\ref{sim:K4}. Number of replications identifying the correct number of significant clusters, median computation time over 100 replications.}
	\label{supptable:K4results4}
\end{table}

\end{document}